\documentclass[12pt]{article}

\newcommand{\VersionInformation}{}  % overwritten by Debug.tex
\InputIfFileExists{Debug}{}{}

\usepackage[utf8]{inputenc}
\usepackage{amsmath}
\usepackage{amsthm}
\usepackage{amsfonts}          % if you want the fonts
\usepackage{amssymb}           % if you want extra symbols
\usepackage[mathscr]{eucal}
\usepackage{graphicx}
\usepackage{color}
\usepackage[nosort]{cite}
\usepackage{hypbmsec}
\usepackage{fancyvrb}
\usepackage{sepnum}
\usepackage{xspace}
\usepackage{booktabs}
\usepackage{rotating}
\usepackage{multirow}
\usepackage[vcentermath]{youngtab}
\usepackage[isu,bf]{caption}
\newlength{\xtrawidth}
\newlength{\xtraheight}
\usepackage[all]{xy}           % Commutative diagrams

\ifx\NoBackreferences\EMPTYMACRO
  \usepackage{hyperref}
\else
  \usepackage[linktocpage,bookmarks]{hyperref}
\fi

\ifx\DEBUG\EMPTYMACRO
\else
  \usepackage{showlabels}
\fi

\def\clap#1{\hbox to 0pt{\hss#1\hss}}

\def\mathrlap{\mathpalette\mathrlapinternal}

\def\mathrlapinternal#1#2{%
\rlap{$\mathsurround=0pt#1{#2}$}}

\makeatletter
  \def\adots{\mathinner{\mkern2mu\raise\p@\hbox{.}
      \mkern2mu\raise4\p@\hbox{.}\mkern1mu
      \raise7\p@\vbox{\kern7\p@\hbox{.}}\mkern1mu}}
\makeatother

\newcommand{\eqdef}{%
  \mathrel{\lower.1mm
    \hbox{$\stackrel{\lower.424ex\hbox{\scriptsize def}}{=}$}}
}
\newcommand{\Q}{\ensuremath{{\mathbb{Q}}}}
\newcommand{\R}{\ensuremath{{\mathbb{R}}}}

\newcommand{\Z}{\mathbb{Z}}
\newcommand{\CP}{{\ensuremath{\mathop{\null {\mathbb{P}}}\nolimits}}}
\newcommand{\IP}{{\ensuremath{\mathop{\null {\mathbb{P}}}\nolimits}}}

\DeclareMathOperator{\tr}{tr}

\DeclareMathOperator{\rank}{rank}

\DeclareMathOperator{\Sym}{Sym}

\DeclareMathOperator{\coker}{coker}

\DeclareMathOperator{\diag}{diag}

\newcommand{\Rep}[1]{\ensuremath{\mathbf{\underline{#1}}}}
\newcommand{\barRep}[1]{\ensuremath{\overline{\Rep{#1}}}}

\DeclareMathOperator{\Ad}{ad}

\newcommand{\Osheaf}{\ensuremath{\mathscr{O}}}

\newcommand{\dP}[1]{\ensuremath{dP_{#1}}}

\newcommand{\Etilde}{\ensuremath{\widetilde{E}}}

\newcommand{\ESixAffineLabels}[7]{
  \ensuremath{
    \vcenter{\xymatrix@C=8mm@R=3mm@!0{
        & & & #4 \ar@{-}[dl] & #2 \ar@{-}[l] \\
        #1 \ar@{-}[r] & #3 \ar@{-}[r] & #5 \\
        & & & #6 \ar@{-}[ul] & #7 \ar@{-}[l] 
      }}
  }
}

\newcommand{\smallESixAffineLabels}[7]{
  \tiny
  \ensuremath{
    \vcenter{\xymatrix@C=5mm@R=.8mm@!0{
        & & & #4 \ar@{-}[dl] & #2 \ar@{-}[l] \\
        #1 \ar@{-}[r] & #3 \ar@{-}[r] & #5 \\
        & & & #6 \ar@{-}[ul] & #7 \ar@{-}[l] 
      }}
  }
}

\newcommand{\IIA}{\ensuremath{\text{II\!A}}}
\newcommand{\IIB}{\ensuremath{\text{IIB}}}

{\end{list}}

{\everymath{\displaystyle\everymath{}}\array}%
{\endarray}

\input cyracc.def %sha
\newfont{\twelvecyr}{wncyr10 at 12pt}
\def\sha{\text{\twelvecyr\cyracc{Sh}}}

\usepackage{color}
\usepackage{listings} 
\usepackage{courier}

\definecolor{dbluecolor}{rgb}{0.01,0.02,0.7}
\definecolor{dgreencolor}{rgb}{0.2,0.4,0.0}
\definecolor{dgraycolor}{rgb}{0.30,0.3,0.30}

\lstdefinelanguage{SageInputLanguage}{
  language=Python, 
  morekeywords={False,sage,True},
  sensitive=true,
}

\lstdefinestyle{SageInput}{
  language=SageInputLanguage,
  basicstyle=\fontsize{11pt}{11pt}\ttfamily\bfseries,
  commentstyle={\ttfamily\color{dgreencolor}},
  stringstyle={\color{dgraycolor}\bfseries},
  keywordstyle=\ttfamily\color{dbluecolor}\bfseries\color{red},
  xleftmargin=25pt,
  belowskip=3pt,
}

\lstdefinelanguage{SageOutputLanguage}{
  morekeywords={False,True},
  sensitive=true,
}

\lstdefinestyle{SageOutput}{
  language=SageOutputLanguage,
  basicstyle={\fontsize{11pt}{11pt}\ttfamily},
  commentstyle={\ttfamily\color{dgreencolor}},
  keywordstyle={\ttfamily\color{dbluecolor}},
  stringstyle={\ttfamily\color{dgraycolor}},
  xleftmargin=25pt,
  aboveskip=0pt,
}

\lstdefinestyle{DefaultSageInputOutput}{
  identifierstyle=,
  numbersep=5pt,
  aboveskip=0pt,
  belowskip=0pt,
  breaklines=true,
  numberstyle=\footnotesize,
  numbers=right,
}

\newcommand{\textsage}[1]{\lstinline[style=SageInput]{#1}}

\begin{document}
%%%%%%%%%%%%%%%%%%%%%[ Title Page ]%%%%%%%%%%%%%%%%%%%%%%%%%%
\begin{titlepage}
  \vspace*{-2cm}
  \VersionInformation
  \hfill
  \parbox[c]{5cm}{
    \begin{flushright}
%      arXiv:yymm.nnnn [hep-th]
    \end{flushright}
  }
  \vspace*{2cm}
  \begin{center}
    \Huge 
    F-theory on Genus-One Fibrations
  \end{center}
  \vspace*{8mm}
  \begin{center}
    \begin{minipage}{\textwidth}
      \begin{center}
        \sc 
        Volker Braun and David R. Morrison
      \end{center}
      \begin{center}
        \textit{
          University of Oxford, Andrew Wiles Building\\
          Radcliffe Observatory Quarter, Woodstock Road\\
          Oxford, OX2 6GG, United Kingdom
        }
      \end{center}
      \begin{center}
        \textit{
          Departments of Mathematics and Physics\\
          University of California, Santa Barbara\\
          Santa Barbara, CA 93106, USA
        }
      \end{center}
    \end{minipage}
  \end{center}
  \vspace*{\stretch1}
  \begin{abstract}
    We argue that M-theory compactified on an arbitrary genus-one fibration, 
    that is, an elliptic fibration which need not have a section, always
    has an F-theory limit when the area of the genus-one fiber approaches zero.
    Such genus-one fibrations can be easily constructed as
    toric hypersurfaces, and various $SU(5)\times U(1)^n$ and $E_6$
    models are presented as examples. To each genus-one fibration one
    can associate a $\tau$-function on the base as well as an
    $SL(2,\Z)$ representation which together define
    the IIB axio-dilaton and $7$-brane content of the theory. The set of
    genus-one fibrations with the same $\tau$-function and $SL(2,\Z)$
    representation, known as the
    Tate-Shafarevich group, supplies an important degree of freedom in the
    corresponding F-theory model which has not been studied carefully until now.

    Six-dimensional anomaly cancellation as well as Witten's zero-mode
    count on wrapped branes both imply corrections to the usual
    F-theory dictionary for some of these models. 
    In particular, neutral hypermultiplets which are
    localized at codimension-two fibers can arise. (All previous known examples
    of localized hypermultiplets were charged under the gauge group of the
    theory.)  
    Finally, in the absence of a section some
    novel monodromies of Kodaira fibers are allowed which lead to new breaking 
    patterns of non-Abelian gauge groups.
  \end{abstract}
  \vspace*{\stretch1}
  \begin{minipage}{\textwidth}
    \underline{\hspace{5cm}}\\
    Email: \texttt{volker.braun@maths.ox.ac.uk}, \texttt{drm@math.ucsb.edu}
  \end{minipage}
\end{titlepage}
\tableofcontents
\listoffigures 	% to produce list of figures
\listoftables 	% to produce list of tables

%%% Local Variables:
%%% TeX-master: "Main"
%%% eval: (TeX-PDF-mode 1)
%%% End:

\section{Introduction}
\label{sec:introduction}

The original F-theory construction~\cite{Vafa:1996xn, Morrison:1996na,
  Morrison:1996pp} produces a non-perturbative type \IIB\ string vacuum
by using a multi-valued function ``$\tau$'' with an $SL(2,\Z)$
symmetry to specify the behavior of the two scalar fields in type \IIB\
string theory.  This function $\tau$ is defined on the complement of a
subvariety $\Delta$ in a compact complex manifold $B$, the base of the
F-theory fibration. Moreover, the $\tau$-function has a specific
asymptotic behavior near the components of the discriminant $\Delta$,
measured by an $SL(2,\Z)$-valued representation of the fundamental group
of $B-\Delta$.  The
components of $\Delta$
serve as sources for the Ramond-Ramond scalar field, i.e., as $7$-branes
in the theory.

The basic F-theory construction comes along with \emph{F-theory/M-theory
duality:}\/ if F-theory is further compactified on a circle, the resulting
theory should be dual to M-theory compactified on a Calabi-Yau variety $Y$
which is fibered over $B$ by curves of genus one, with the curves becoming
singular over $\Delta$, and with the function $b\mapsto \tau(b)$
corresponding to the
ratio of periods of the holomorphic $1$-form on the fiber over $b$.  
As the area of the genus-one curves shrinks,
the dual circle expands; the zero-area limit corresponds to 
decompactification to the original F-theory model.  This duality arises
by a fiberwise application of an M-theory/Type IIB duality 
\cite{Schwarz:1995dk,Aspinwall:1995fw}.

The Calabi-Yau varieties $Y$ whose M-theory compactifications have
F-theory limits should therefore admit genus-one fibrations $Y\to B$,
i.e., maps $\pi:Y\to B$ whose general fiber is a curve of genus one.
Traditionally in the F-theory literature one has demanded that $\pi:Y\to B$
have a section but we relax that requirement in this paper.
We do, however, require that the fibration $\pi$ is flat, i.e., that all
fibers of $\pi$ are one-dimensional.
(Otherwise, tensionless strings will give additional
massless degrees of freedom in addition to the desired low-energy
effective Yang-Mills theory.)

Since $Y$ is an algebraic variety whenever $\dim Y\ge3$, 
it must have an embedding into
projective space with some equations defining the image; the existence of
the genus-one fibration means that those equations define a family
of curves of genus one parametrized by $B$.  In particular, these equations
can be regarded as defining a curve of genus one whose defining equations
have coefficients which are locally defined functions on $B$.  By
allowing denominators in the coefficients, we can instead think of
this as a curve of genus one whose defining equations have coefficients
 chosen from the field $K$ of rational functions on $B$.  The equations
for various open subsets of $B$ will all induce the same equation
with coefficients in $K$.

Algebraic geometers and number theorists have long studied algebraic curves
whose defining equations have coefficients in
a field $K$ that may not be algebraically closed.  If a curve of genus
one over $K$ has a point on it with coefficients in $K$, then the curve
is called an \emph{elliptic curve over $K$}\/ 
and its points form an Abelian group.
When $K$ is the function field of a base manifold $B$, the points with
coefficients in $K$ correspond to \emph{rational sections}\/ of the
genus-one fibration; thus, a genus-one fibration is an elliptic fibration
exactly when there is a rational section.

Every algebraic curve $C$ of genus one over $K$ has an associated
elliptic curve $J(C)$ over $K$ known as the \emph{Jacobian}\/ of $C$
as we will review in \autoref{sec:Jacobian}. Geometrically, the points
on the Jacobian represent line bundles of degree zero on $C$, and
since there is a distinguished line bundle (the trivial line bundle)
the Jacobian has a distinguished point.  Moreover, since the set of
line bundles of degree zero forms a group (under tensor product), the
group structure on $J(C)$ is evident.

If $Y$ is a Calabi-Yau surface or threefold with a genus-one
fibration, then the (compactified)
Jacobian of the fibration determines an associated
Calabi-Yau variety\footnote{This statement is presumably also true for
  Calabi-Yau fourfolds, but a mathematical proof has not yet been
  given.} $J(Y)$ with an elliptic fibration
\cite{MR0184257,MR977771}. Moreover, the multi-valued $\tau$ functions 
and $SL(2,\Z)$ representations for $Y\to
B$ and $J(Y)\to B$ are identical.  (Even the discriminant loci for the
fibrations are identical as subvarieties of $B$ \cite{MR1242006}.)
For this reason, one should expect
that the F-theory limits of M-theory compactified on $Y$ and on $J(Y)$ are 
identical.  It has
often been asserted---particularly by the second author of this
paper---that this fact makes the study of non-elliptic genus-one
fibrations superfluous for F-theory.  However, as we will show in this
paper, the compactified Jacobian $J(Y)$ typically has singularities for a
nonsingular genus-one fibered Calabi--Yau threefold $Y$ without a
section.  The singularities in question  are ``$\Q$-factorial terminal''
singularities (the simplest example being a conifold with no small
resolution) which have the property that any blowup of them introduces zeros 
into the holomorphic $3$-form.

The presence of such singularities on $J(Y)$ at first leads one to expect
that M-theory cannot be compactified on this space.  However, since
the nonsingular space $Y$ has a good F-theory limit and shares the same
$\tau$-function and monodromy representation with $J(Y)$, such an interpretation
 would be rather
puzzling.  To resolve this puzzle, one should remember that when the
M-theory $3$-form has torsion flux, there can be so-called
``frozen'' singularities
linked to that flux which are not resolved in the M-theory compactification,
and which do not contribute to the M-theory gauge group in the standard
way \cite{Witten:1997bs,triples}.
Although a complete understanding of such torsion fluxes awaits a
better understanding of the M-theory $3$-form, we shall argue
in this paper that this is indeed the correct explanation.

In practice, it is often easier to study $Y$ than $J(Y)$, so these
genus-one fibrations are \emph{not}\/ superfluous.
In fact, as we will see in detail in \autoref{sec:hyper_count}, 
and as had been earlier observed in \cite{Bouchard:2003bu}, this
phenomenon already happens for Calabi-Yau threefold hypersurfaces in
rather simple toric varieties. Still, one might ask how F-theory on
genus-one fibrations is really different from the familiar case of
elliptic fibrations. In this paper, we will point out the following
two:
\begin{itemize}
\item There can be massless fields localized on codimension-two
  fibers. In fact, subtleties in the interpretation of the F-theory
  limit for a genus-one fibration without a section were observed
  already in~\cite{Berglund:1998rq} which considered a ``degree two''
  case in the language of this paper. In \autoref{sec:anomaly} we will
  explain the origin of these massless fields. In particular, 
the existence of such fields
  corrects the hypermultiplet moduli count of 6-d F-theory
  compactifications. After taking this correction into account, the
  gravitational anomaly is again canceled.
\item There are more possibilities for the monodromies of the Kodaira
  fiber over a discriminant component. In fact, if one requires a
  section then one of the $\CP^1$-components with multiplicity 1
  (corresponding to a node of the Dynkin diagram with Dynkin label 1)
  must be fixed by the monodromy. This need not be the case if there
  is no section, yielding new ways to break non-Abelian gauge symmetry
  by monodromies. For example, in \autoref{sec:E6} we will discuss an
  example of a toric hypersurface where monodromy breaks $E_6\to
  G_2$. The Shioda-Tate(-Wazir) formula for the cohomology groups of
  elliptic fibrations naturally generalizes to genus-one fibrations
  with more general monodromies, see \autoref{sec:STW}.
\end{itemize}
On the base $B$ of the elliptic
fibration, the data defining the genus-one fibration $Y\to B$ is the
$\tau$-function and $SL(2,\Z)$ representation (or equivalently, the 
compactified Jacobian $J(Y)$), together with a
class in the \emph{Tate-Shafarevich} group \cite{MR0106226,MR0162806}
of isomorphism classes of elliptic fibrations with the same Jacobian.
The Tate-Shafarevich group $\sha_B(J(Y))$ can be described in in terms of certain
sheaf cohomology groups on the base $B$ (which make it obvious that
it is indeed a group). The simplest description \cite{MR1242006,MR1918053,MR2399730},
which holds when $J(Y)$ and $B$ are both smooth, all fibers of the fibration
are irreducible,\footnote{Note that this condition implies no non-Abelian
gauge symmetry in the F-theory model; the description must be modified
in more general cases.}
and the section of $J(Y)\to B$ is regular, is
\begin{equation}
  \sha_B(J(Y)) = H^1(B, \mathcal{A}),
\end{equation}
where $\mathcal{A}$ is the sheaf of rational sections of $J(Y)\to B$.
The interpretation in F-theory is of a degree of freedom which can
be given an expectation value on a loop when compactifying F-theory to
M-theory, resulting in a collection of M-theory vacua with the same
F-theory limit \cite{wilson-lines, triples}.

\section{From Genus-One Fibrations to Elliptic Fibrations}
\label{sec:Jacobian}

Let us start by reviewing how to assign the multi-valued $\tau$
function to a genus-one fibration, that is, how the defining equations
of the given fibration determine defining equations of the Jacobian
fibration. (The latter can always
be brought into Weierstrass form.) A Calabi-Yau variety $Y$ of
dimension $k$ with a genus-one fibration always has an embedding into
projective space when $k\ge3$.\footnote{The case $k=2$ is special
  because a K3 surface may fail to be a projective variety.}  Let us
choose a line bundle $\mathcal{L}$ on $Y$ which is ample on the general fiber
of the fibration,\footnote{That is, the sections of
  some power of $\mathcal{L}$ give a map to projective
  space which is an embedding of the general fiber.} and let $d$ be the degree of
$\mathcal{L}$ when restricted to the curves of genus one.  That is, the
associated curve $C$ of genus one over the function field $K$ has a
line bundle $\mathcal{L_C}$ of degree $d$. For small values of $d$,
the properties of $\mathcal{L_C}$ determine the structure of a
birational model $\overline{Y}$ of $Y$.  This determination is based
on the Riemann-Roch theorem for $C$, which continues to hold even over
a non-algebraically closed field.  For a line bundle $\mathcal{M}$ of
positive degree on a curve of genus one, $H^1(C,\mathcal{M})\cong
H^0(C,K_C\otimes \mathcal{M}^{-1})^*$ which vanishes since
$\deg(K_C\otimes \mathcal{M}^{-1})=0-\deg(\mathcal{M})<0$.  Thus,
\begin{equation}
  \dim H^0(C,\mathcal{M}) = \chi(C,\mathcal{M}) = \deg(\mathcal{M})-g+1
  =\deg(\mathcal{M})  
\end{equation}
since $g=1$.

If $C$ has a line bundle $\mathcal{L}$ of degree $1$, this leads to
``Weierstrass form'' in the following way, as explained by
Deligne~\cite{MR0387292} (see also \cite{MR0437531,MR977771}).  $H^0(C,\mathcal{L})$ has dimension one, so
choose a basis element $z$ for this space.
$H^0(C,\mathcal{L}^{\otimes 2})$ has dimension two and contains $z^2$;
choose another element $x$ so that $z^2, x$ is a basis of this space.
$H^0(C,\mathcal{L}^{\otimes 3})$ has dimension three and contains
$z^3$ and $xz$; choose another element $y$ so that $z^3, xz, y$ is a
basis of this space.  We can now enumerate the sections we know of
various powers of $\mathcal{L}$:
\begin{equation}
  \begin{split}
    H^0(C,\mathcal{L})\phantom{{}^{\otimes 1}} & \ni z \\
    H^0(C,\mathcal{L}^{\otimes 2}) & \ni z^2, x \\
    H^0(C,\mathcal{L}^{\otimes 3}) & \ni z^3, xz, y \\
    H^0(C,\mathcal{L}^{\otimes 4}) & \ni z^4, xz^2, yz, x^2 \\
    H^0(C,\mathcal{L}^{\otimes 5}) & \ni z^5, xz^3, yz^2, x^2z, xy\\
    H^0(C,\mathcal{L}^{\otimes 6}) & \ni z^6, xz^4, yz^3, x^2z^2, 
    xyz, x^3, y^2.
  \end{split}
\end{equation}
Since there are seven sections of $H^0(C,\mathcal{L}^{\otimes 6})$,
there must be a relation among them.  As Deligne argues, since there is
no relation among sections of
 lower powers of $\mathcal{L}$, the coefficients of $x^3$
and $y^2$ in the relation must be nonvanishing, and after rescaling we
can assume those coefficients are $1$.  Thus, we get an equation of
the form
\begin{equation}
  y^2 + a_1 xyz + a_3 yz^3 = x^3 + a_2 x^2z^2 + a_4 xz^4 + a_6 z^6,
\end{equation}
the long Weierstrass from. If the characteristic of $K$ is not $2$ or
$3$, as is true in our case, we arrive at an equation of the form
\begin{equation}
  y^2 = x^3 + fx^2z^4 + gz^6.
\end{equation}
by completing the square on $y$ and completing the cube on $x$. This
is the Weierstrass form.

We can ``clear denominators'' in the coefficients $f$, $g\in K$ in the
Weierstrass equation and take them to lie in the ring of functions on
any subset of the base $B$.  More generally, $f$ and $g$ can be taken
as sections of appropriate line bundles over $B$, and this gives the
most general Weierstrass forms of elliptic fibrations. Note that any
genus-one fibration of degree $1$ has a section (given by $z=0$ in
the coordinates above); that is, it is an elliptic fibration.

There are similar stories for other low values of the degree $d$. 
First, one can analyze sections of the line bundle to determine
the form of the equation of the curve $C$ of genus one.  Then, the
geometric construction of the Jacobian fibration can be studied
algebraically, resulting in a formula for the equation of the
Jacobian $J(C)$,
given the equations for $C$.  For degrees $2$, $3$, and $4$, such
formulas are derived and presented in a systematic way
in~\cite{MR1858080}.\footnote{There is an ``improved version'' of
  these formulas in \cite{MR2183258} which work well in characteristic
  $2$ and $3$.}  Weil~\cite{MR0061857} has traced the history back to
Hermite. The formulas also appear in Duistermaat's
book~\cite{pre05771645}.

When $d=2$, we argue as follows.\footnote{This argument is well-known
in the mathematics literature, and was written out in Appendix~B 
of~\cite{MW}.}  This time, let $\mathcal{M}$ be a line bundle on $C$ of
degree $2$.  We choose a basis $u$, $v$ of the two-dimensional vector
space $H^0(C,\mathcal{M})$, and then choose an element $w$ of
$H^0(C,\mathcal{M}^{\otimes 2})$ so that $u^2, uv, v^2, w$ is a basis
of that space. Once again, we enumerate the sections we know of
various powers:
\begin{equation}
  \begin{split}
    H^0(C,\mathcal{M})\phantom{{}^{\otimes 1}} & \ni  u, v  \\
    H^0(C,\mathcal{M}^{\otimes 2}) & \ni u^2, uv, v^2, w \\
    H^0(C,\mathcal{M}^{\otimes 3}) & \ni u^3, u^2v, vv^2, v^3, uw, 
    vw \\
    H^0(C,\mathcal{M}^{\otimes 4}) & \ni u^4, u^3v, u^2v^2, uv^3, 
    v^4, u^2w, uvw, v^2w, w^2.\\     
  \end{split}
\end{equation}
Since there are nine sections of $H^0(C,\mathcal{M}^{\otimes 4})$,
there must be one relation among them.  Again, since there is no
relation among sections of lower powers of $\mathcal{M}$, the
coefficient of $w^2$ in the relation must be nonvanishing, and after
rescaling we can assume that coefficient is $1$. Thus, we get an
equation of the form
\begin{equation}
  w^2 + b_0 u^2w + b_1uvw + b_2 w^2 = c_0 u^4+c_1u^3v + c_2u^2v^2 + c_3uv^3
  +c_4v^4.
\end{equation}
If the characteristic of $K$ is not $2$, which we always assume in
this paper, we can simplify this to an equation of the form
\begin{equation} \label{eq112}
  w^2 = e_0 u^4 + e_1 u^3v + e_2 u^2v^2 + e_3 uv^3 + e_4v^4.
\end{equation}
by completing the square on $w$. This is standard form for degree two.

We will give a derivation of the formula for the Jacobian of $C$
in degree $2$, based on
some Galois theory which is often found in undergraduate abstract
algebra classes.  We start with a polynomial of degree $4$ in a single
variable, obtained from the right hand side of eq.~\eqref{eq112} by
setting $v=1$:
\begin{equation}
  e_0 u^4 + e_1 u^3 + e_2 u^2 + e_3 u + e_4= 
  e_0\left( u^4 + \frac{e_1}{e_0} u^3 + \frac{e_2}{e_0} u^2 + \frac{e_3}{e_0}u
    +\frac{e_4}{e_0}\right) .
\end{equation}
Galois theory constructs a finite extension $L$ of the field $K$ in
which this polynomial has roots, and as is well-known (and visible
in the classic
formulas of Cardano), the first step is to define an associated polynomial
of degree $3$.  In fact, over the field $L$ the polynomial will factor as
\begin{equation}
  e_0\prod_{i=1}^4(u-r_i),
\end{equation}
and the cubic polynomial is related to this by forming the quantities
\begin{equation}
  s_1=r_1r_2+r_3r_4, \quad s_2=r_1r_3+r_2r_4, \quad s_3=r_1r_4+r_2r_3,
\end{equation}
and using those to form the auxiliary cubic polynomial
\begin{equation}
  \prod_{j=1}^3 (\widetilde{x}-s_j). 
\end{equation}
Note that
\begin{equation}
  \begin{aligned} 
    s_1-s_2&=(r_1-r_4)(r_2-r_3)\\
    s_1-s_3&=(r_1-r_3)(r_2-r_4)\\
    s_2-s_3&=(r_1-r_2)(r_3-r_4)\\
  \end{aligned}
\end{equation}
which implies that the discriminants of the two polynomials are the same.

We can determine the equation of the auxiliary cubic polynomial by 
expressing the elementary symmetric functions of its roots in terms of
the elementary symmetric functions $\sigma_1$, \dots, $\sigma_4$
of $\{r_1,\dots,r_4\}$.  The calculation is straightforward, and the result
is as follows:
\begin{equation}
  \begin{aligned}
    s_1 + s_2 + s_3 &= \sigma_2 \\
    s_1 s_2 + s_1 s_3 + s_2s_3 &= \sigma_1\sigma_3-4\sigma_4\\
    s_1s_2s_3 &= \sigma_1^2\sigma_4+\sigma_3^2-4\sigma_2\sigma_4\\
  \end{aligned}
\end{equation}
This implies that the cubic polynomial takes the form
\begin{equation}
  \widetilde{x}^3 - \frac{e_2}{e_0} \widetilde{x}^2
  + \frac{e_1e_3-4e_0e_4}{e_0^2} \widetilde{x}
  -\frac{e_1^2e_4+e_0e_3^2-4e_0e_2e_4}{e_0^3}.
\end{equation}
If we rescale by substituting $x=e_0\widetilde{x}$ we find a cubic polynomial
\begin{equation}
  x^3 - e_2 x^2
  + (e_1e_3-4e_0e_4)x
  -(e_1^2e_4+e_0e_3^2-4e_0e_2e_4).
\end{equation}
The corresponding homogenous equation
\begin{equation}
  y^2 = x^3 - e_2 x^2z^2
  + (e_1e_3-4e_0e_4)xz^4
  -(e_1^2e_4+e_0e_3^2-4e_0e_2e_4)z^6
\end{equation}
is the equation of the Jacobian of the curve with equation \eqref{eq112}.
(To put it into Weierstrass form one should complete the cube in $x$.)

The Galois theory which goes along with this construction is based on
the following exact sequence of groups:
\begin{equation}
  1 \to (\mathbb{Z}_2)^2 \to \mathfrak{S}_4 \to \mathfrak{S}_3 \to 1.
\end{equation}
As a consequence, 
if $L$ is the field in which  the degree $4$ polynomial (assumed to be general)
has its
roots so that the Galois group of $L$ over $K$ is the symmetric group 
$\mathfrak{S}_4$, there is
an intermediate field $L'$ whose Galois group over $K$ is 
$\mathfrak{S}_3$.  The above
construction explicitly builds the degree $3$ polynomial whose roots lie in the
field $L'$.

Returning to the general problem of a curve of genus one over $K$ with a
line bundle of degree $d$, if $d=3$ the story is straighforward: let
$\mathcal{N}$ be a line bundle of degree $3$, and choose a basis $x$,
$y$, $z$ of $H^0(C,\mathcal{N})$.  There are $6$ degree two monomials
in $x$, $y$ and $z$ which matches the dimension of
$H^0(C,\mathcal{N}^{\otimes 2})$.  However, since there are $10$
degree three monomials in $x$, $y$, and $z$ but the space
$H^0(C,\mathcal{N}^{\otimes 3})$ has dimension $9$, there must be a
relation of degree $3$.  This expresses $C$ as a cubic curve in
$\mathbb{P}^2_K$. The formula for the Weierstrass equation of the
Jacobian is quite lengthy in this case, but can be found in \cite{MR2183258}.

If $d=4$ the story is also straightforward: let $\mathcal{P}$
be a line bundle of degree $4$, and choose a basis $x$, $y$, $z$, $t$
of $H^0(C,\mathcal{P})$.  There are $10$ degree $2$ monomials in
$x$, $y$, $z$, and $t$ yet the space $H^0(C,\mathcal{P}^{\otimes 2})$
has dimension $8$, so there must be two relations of degree $2$.  Let
$Q_1$, and $Q_2$ be a basis of the space of relations; then $C$ is described
as a complete intersection $\{Q_1=Q_2=0\}$ in $\mathbb{P}^4_K$.

To describe the Weierstrass equation of the Jacobian in this case, we
represent $Q_1$ and $Q_2$ as symmetric $4\times 4$ matrices, and
consider the determinant $\det(\lambda Q_1 + \mu Q_2)$.  This is a
homogeneous polynomial of degree $4$ in $\lambda$ and $\mu$, and
thereby determines a curve $C'$ of genus one and degree $2$ as the double
cover of $\mathbb{P}^1_K$ branched on the zeros of that homogeneous
polynomial. The Jacobian of $C$ coincides with the Jacobian of $C'$,
and our previous formula for the degree $2$ case provides the
Weierstrass equation of the Jacobian, an explicit formula for which can be found
in~\cite{MR0094753, MR1858080, trac14855}. If the coordinates $x$,
$y$, $z$, $t$ can be chosen such that one of the quadratics, say,
$Q_1$, takes the special form
\begin{equation}
  Q_1 = xy -zt,
\end{equation}
then $\{Q_1=0\}$ is itself a toric variety $\mathbb{P}^1_K\times
\mathbb{P}^1_K$ and we again have a hypersurface in a toric variety.

The story above has recently been extended to $d=5$.
There are $5$ sections of the defining line bundle, and the equations can
be presented as the vanishing of the
$4\times4$ Pfaffians of a $5\times5$ matrix
with entries that are linear functions of the $5$ sections
\cite{buchsbaum1977algebra}.  Formulas have been found for the Weierstrass equation
of the Jacobian of the curve, and although the formulas
are too large to write down as explicit polynomials, there is an algorithm
for evaluating them \cite{MR2448246}.

\section{Singularities of Jacobian Fibrations}
\label{sec:singularities}

Dolgachev and Gross \cite{MR1242006} have studied
 the Tate-Shafarevich group  of a genus-one fibered
threefold in considerable detail.  Their analysis goes beyond
the Calabi-Yau case and includes arbitrary genus-one fibered algebraic
threefolds.  The results fall short of giving an algorithm for
computing the Tate-Shafarevich group, but they are strong enough to compute
it in some important examples.  The analysis is quite technical
and we will not attempt to present the results here.
However, many of the key features of their analysis are present
in one particular example which we now describe following \cite{MR1242006}.

Let $f_1(\vec{x})$, $f_2(\vec{x})$, $f_3(\vec{x})$ be three general cubics in $\mathbb{P}^2$, and consider
\begin{equation}
Y = \{(\vec{x},\vec{u})\in \mathbb{P}_{\vec{x}}^2\times \mathbb{P}_{\vec{u}}^2\ |\
u_1\,f_1(\vec{x}) + u_2\,f_2(\vec{x}) + u_3\,f_3(\vec{x})=0\}.
\end{equation}
We map $Y\to \mathbb{P}_{\vec{u}}^2$ by $(\vec{x},\vec{u})\mapsto \vec{u}$, and
note that the fiber over $\vec{u}$ is a curve of genus one in $\mathbb{P}^2_{\vec{x}}$.  This fibration has no section if the three cubics are general,
but there is a 3-section given by intersecting $Y$ with $x_3=0$, which
gives three points in each fiber.
(We stress  that this example is {\em not} Calabi-Yau, but is nevertheless
fibered  by curves of genus one.)

The discriminant locus for $Y\to\mathbb{P}_{\vec{u}}^2$ can be computed with some
methods from classical algebraic geometry.  The computation was made
in \cite{MR644816}, and  the result states that the discriminant is an irreducible
 curve in
$\mathbb{P}^2_{\vec{u}}$ of degree $12$ with $24$ cusps and $21$ nodes.
Moreover, the total space $Y$ of this family is nonsingular, and the map $Y\to \mathbb{P}^2_{\vec{u}}$ is flat; this enables Dolgachev and Gross to identify
the Tate-Shafarevich group in this case
with a subgroup of $\mathbb{Z}/3\mathbb{Z}$, using
their general results about Tate-Shafarevich groups for flat
fibrations with nonsingular total space.  Since $Y\to\mathbb{P}^2_{\vec{u}}$ itself
does not have a section, the Tate-Shafarevich group must be nontrivial,
so it must be $\mathbb{Z}/3\mathbb{Z}$.

A key result of \cite{MR1242006} is that the Jacobian fibration 
$J(Y)\to\mathbb{P}^2_{\vec{u}}$ has the same discriminant locus as that of $Y\to \mathbb{P}^2_{\vec{u}}$.
Thus, the Jacobian fibration in this example
has a Weierstrass model whose discriminant is
a curve of degree $12$ with $24$ cusps and $21$ nodes.  Now over a general
point of the discriminant locus in a Weierstrass fibration, the fiber
acquires a node (Kodaira type $I_1$) but the total space is smooth.  Similarly,
at a cusp in the discrimiant of a Weiestrass model, the fiber has a
cusp (Kodaira type $II$) but the total space is smooth.  The only
place where the total space of the Weierstrass model is singular is
at the $21$ nodes of the discriminant locus, where the Kodaira type of
the fiber is $I_2$. The singularity in the total space is an ordinary
quadratic singularity, otherwise known as a ``conifold'' 
singularity.\footnote{Note that all of these statements could be explicitly
verified using the techniques of Section~\ref{sec:Jacobian}.}

In order to produce a nonsingular model of $J(Y)\to \mathbb{P}^2_{\vec{u}}$, one
would like to find a ``small resolution'' of the conifold singularity.  The standard way to
do this would exploit the local factored form of the discriminant:
the singularity can locally be written in the form $xy=h(u_1,u_2)$ 
in the affine chart $u_3=1$ with
$h(u_1,u_2)$ representing the discriminant, and taking a local factorization
$h=h_1h_2$ (where $h_j=0$ defines one of the local branches of the curve
at its node, for $j=1, 2$),
the blowup of $\{x=h_1=0\}$ produces the small resolution.
Such a small resolution, if it existed, 
 would give a flat family $\widetilde{J(Y)}\to\mathbb{P}^2_{\vec{u}}$ with
nonsingular total space.  However, since $h(u_1,u_2)$ is globally an
irreducible curve, this factorization cannot be performed globally and
the small resolution may not exist as an algebraic variety.\footnote{The
fact that ``collisions'' between Kodaira fibers of type $I_{2k+1}$ and
$I_{2\ell+1}$ may obstruct the existence of flat families was first observed by
Miranda \cite{MR690264}.}  The failure of small resolutions to exist globally
has shown up a number of times in the past in applications to string theory,
including \cite{dP8}, where it was responsible for a St\"uckelberg mechanism
in effective theories of D-branes at singularities.

In fact, as shown in \cite{MR1242006}, there is {\em no} small resolution of
the conifold singularities in this example.  If there were, then the
general results of \cite{MR1242006} would imply that the Tate-Shafarevich group of $J(Y)$
is a subgroup of $\mathbb{Z}/1\mathbb{Z}$, i.e., it would have to be the trivial
group.\footnote{According to \cite{MR1242006},
the Tate-Shafarevich group
for a flat family with nonsingular total space is an extension of
a certain geometric group by $\mathbb{Z}/\delta \mathbb{Z}$, where $\delta$ is the minimum intersection number of a fiber with 
a multisection.  For $Y$, $\delta=3$, but for $J(Y)$, $\delta=1$.}
But since the Tate-Shafarevich group is actually $\mathbb{Z}/3\mathbb{Z}$, the
small resolution cannot exist.

Notice that it is always possible to blowup the conifold point itself, giving
a ``big'' resolution.  However, any holomorphic three-form on $J(Y)$ which
is nonvanishing near the conifold point necessarily has zeros along the
exceptional divisor of the big resolution.  Therefore, the big resolution
is not suitable for studying Calabi-Yau threefolds.  It also fails to
have a flat fibration.

The conclusion that the Jacobian of a nonsingular genus-one fibration has
conifold singularities (or possibly worse singularities) which cannot
be resolved to give a flat family appears to be a general one, borne out
by additional examples in \cite{MR1242006} as well as further examples
in this paper.  The general theory of minimal models in the birational
geometry of threefolds (see \cite{Kawamata:mmp}, for example) identifies these singularities
as ``$\mathbb{Q}$-factorial terminal singularities.''

In addition to conifold singularities which cannot be
resolved, Dolgachev and Gross find another geometric feature of $J(Y)$ 
in this example which
(as we will see) helps to identify the physics of the corresponding
M-theory compactification.  
Let us blow up the base at the nodes of the discriminant, and then
blowup the singular locus of the total space, to obtain
a new surface $\widehat{\mathbb{P}^2}$ with an elliptic fibration
$\widehat{J(Y)}\to \widehat{\mathbb{P}^2}$ which is flat and has a nonsingular
total space $\widehat{J(Y)}$.  The Tate-Shafarevich group of $\widehat{J(Y)}$
is again $\mathbb{Z}/3\mathbb{Z}$, and this time can be identified with
the torsion in cohomology, i.e., with\footnote{This identification
proceeds \cite{MR1242006,MR2399730} via the cohomological Brauer group $Br'(\widehat{J(Y)})=H^2(\widehat{J(Y)},\mathcal{O}_{\widehat{J(Y)}}^\times)$,
which coincides with the torsion in cohomology since $\widehat{J(Y)}$
is nonsingular \cite{brauer}, a result which holds since $\widehat{J(Y)}$
has no holomorphic $1$-forms or $2$-forms.}
 $H^3(\widehat{J(Y)},\mathbb{Z})_{\text{tors}}$.
Note that since the torsion is a birational invariant, this can also
be seen as a torsion cohomology class on the ``big blowup'' of the
conifold points on $J(Y)$.

As observed in \cite{Witten:1997bs} and studied further in \cite{triples},
there are M-theory compactifications on certain singular spaces (spaces with
``frozen'' singularities) which
are well-defined in spite of the singularities; the interpretation is that
the M-theory $3$-form has a discrete flux which obstructs the resolution 
of the singularities.  We find ourselves in a similar situation here,
with an M-theory model on the singular space $J(Y)$ which (due to having
an F-theory limit in common with M-theory on $Y$) should be well-defined.
Moreover, the space has a natural torsion $3$-form.\footnote{There is a
subtlety here, in that the torsion $3$-form exists not on $J(Y)$ itself
but on a blowup.  However, as shown in \cite{stabs}, in the context of
explaining some examples of discrete torsion in type IIA string theory
\cite{Vafa:1994rv}, a torsion class on a blowup of $J(Y)$ can lead to
corresponding physical effects on $J(Y)$ itself.  The torsion in
\cite{stabs} was related to the cohomological Brauer group 
in that case just as it is in the present case.}

This, then is our interpretation:  the Jacobian of a genus-one fibered
Calabi-Yau threefold $Y$
is an elliptic fibration $J(Y)$ with $\mathbb{Q}$-factorial terminal 
singularities which is equipped with an appropriate torsion class in
such a way that M-theory compactified on $J(Y)$ with torsion flux for
the M-theory $3$-form is well-defined.

%%% Local Variables:
%%% TeX-master: "Main"
%%% mode: TeX-PDF-mode on
%%% End:

\section{Fiberwise Duality with M-Theory}

\subsection{Review of the F/M Duality}

One way to define F-theory compactified on an elliptic fibration is to
compactify M-theory on one of the cycles of the torus fiber down to
\IIA\ and then perform T-duality~\cite{Denef:2008wq} to \IIB. The
T-dual circle decompactifies in the limit where the torus fiber
shrinks to vanishing size, and in fact one obtains a \IIB\ string
theory compactified on the base of the elliptic fibration with varying
axion-dilaton.

The argument goes as follows. Pick coordinates $z=x+\tau y$ with
$0\leq x, y\leq 1$ on the torus $E_\tau$. Then the Calabi-Yau metric
on a $E_\tau$-fibration $Y$ over $B$ is presumably of the form
\cite{MR1059826,GrossWilson:LCSL}
\begin{equation}
  \label{eq:Fmetric}
  ds^2_Y = 
  ds^2_B + 
  \frac{v}{\tau_2}dz
  d\bar{z} +
  O(v^2)
\end{equation}
in the limit where the fiber volume $v\to 0$. The complex structure
$\tau=\tau_1+i\tau_2$ of the fiber $E_\tau$ varies as a function of
the base. One recognizes this as the M-theory lift, on the $x$-circle,
of the \IIA\ metric
\begin{equation}
  \begin{gathered}
    ds^2_M = e^{4\phi_\IIA/3} 
    \big( dx + C_1 \big)^2  + e^{-2\phi_\IIA/3} ds^2_{IIA},
    \\
    C^{(1)}_\IIA = \tau_1 dy
    ,\quad
    e^{4\phi_\IIA/3} = \frac{v}{\tau_2}
    ,\quad
    ds^2_\IIA = \sqrt{\frac{v}{\tau_2}} \big(v \tau_2 dy^2 + ds^2_B
    \big)
    .
  \end{gathered}
\end{equation}
To dualize to type \IIB, we now perform T-duality on the remaining
$y$-circle. Its circumference, in the \IIA\ metric, is $L_y = v^{3/4}
\tau_2^{1/4}$. Hence, T-duality amounts to rescaling
\begin{equation}
  T: y \mapsto v^{-3/2} \tau_2^{-1/2} y
\end{equation}
which results in the \IIB\ metric
\begin{equation}
  ds^2_{\IIB,S} = \sqrt{\frac{v}{\tau_2}} 
  \left( \frac{dy^2}{v^2} + ds^2_B \right)
  .
\end{equation}
Moreover, T-duality shifts the dilaton $\phi$ and identifies
corresponding RR-forms with one leg in the circle direction, namely
\begin{equation}
  e^{\phi_\IIB} = \frac{1}{L_y} e^{\phi_\IIA} = \frac{1}{\tau_2}
  ,\qquad
  C^{(0)}_\IIB dy =
  C^{(1)}_\IIA.
\end{equation}
Finally, we rescale the \IIB\ metric from the string
frame to the Einstein frame to obtain
\begin{equation}
  ds^2_{\IIB,E} = e^{-\frac{\phi_\IIB}{2}} ds^2_{\IIB,S} = 
  \sqrt{v} 
  \left( \frac{dy^2}{v^2} + ds^2_B \right)
  .
\end{equation}
Remarkably, the $\tau$-dependence of the metric has completely
disappeared. The fiber volume $v$ is constant over the base, so the
metric is just a metric on $[0,1] \times B$ and decompactifies to $\R
\times B$ in the limit $v\to 0$. The entire $\tau$-dependence is in
the \IIB\ axion-dilaton, which is simply $C^{(0)}_\IIB+ie^{-\phi_\IIB}
= \tau_1 + i \tau_2 = \tau$

\subsection{Fiberwise Duality}

The argument presented above is clearly na\"{\i}ve: The fiber complex
structure $\tau$ varies holomorphically, so if it is non-constant then
it must have zeros and poles where the ansatz eq.~\eqref{eq:Fmetric}
cannot be valid. Even worse, there are $SL(2,\Z)$ transformations
along loops which encircle
the discriminant locus $\Delta\subset B$, so there is no global choice
of $x$ and $y$-circle.

Hence, we also need to appeal to fiberwise duality to complete the
M-theory/\IIB\ duality. Locally, over the base $B$, there is no
preferred $SL(2,\Z)$ frame. But that choice also has no physical
significance: possible Dehn twists on the F-theory elliptic curve just
correspond to the changing S-duality frame of the \IIB\ axion-dilaton. We can
apply the above duality on sufficiently small open sets and glue the
\IIB\ description via $SL(2,\Z)$ transformations. The result is \IIB\ on
the base with varying axion-dilaton $\tau$. The actual value of $\tau$
is not uniquely defined, but it is unique up to
$SL(2,\Z)$-transformations. The  representation
$\pi_1(B-\Delta)\to SL(2,\Z)$ is part of the defining data of the
elliptic fibration.

However, $SL(2,\Z)$-transformations are not the entire symmetry group
by which one can glue local patches. In addition, there are
translations along the fiber. This obviously does not preserve the
zero-section (i.e., the locus of points serving as ``$0$'' in the
group structure on each fiber), 
so the ensuing fibration will, in general, only be a
genus-one fibration. At first sight, allowing translations seems to
be very boring: $\tau$ does not change if we translate along the
torus, so no physical quantity appears to know about it. However, this
argument really only tells us that no field knows about the
translations locally, which is tautologically true, as the geometry
has local sections. But global monodromies can and will depend on this
additional freedom, and in \autoref{sec:3x_monodromy} we will see an
explicit example.

\subsection{Tate-Shafarevich Group}

To be able to act requires an identity element on the part of the
actor, but not on the part of the acted upon. A vector space acts on
an affine space by translations. Or, relevant for our purposes, an
elliptic curve acts on a curve of genus one with the same $\tau$ by
translations. In particular, recall from \autoref{sec:Jacobian} that
for every genus-one fibration $Y$ there is an elliptic fibration
$J(Y)$. Sections of the Jacobian act by translation on the genus-one
fibration, turning the Mordell-Weil group of the Jacobian into a
subgroup of the birational symmetries of $Y$.

To construct \emph{new} genus-one fibrations, we can start with an
elliptic fibration $Y\to B$ and choose translations locally in each
patch of the base $B$. Gluing together the patches by the translations
creates a genus-one fibration which may or may not have
sections. Explicitly, let $\mathcal{A}$ be the sheaf of rational
sections of the elliptic fibration. By definition, $\mathcal{A}$ is a
sheaf of Abelian groups with respect to fiber-wise addition, which we
write as ``$+$''. We are interested in a collection
$\sigma_{\alpha\beta} \in \mathcal{A}(U_\alpha\cap U_\beta)$ of local
sections that fits together on triple overlaps, that is,
$\sigma_{\alpha\beta} + \sigma_{\beta\gamma} =
\sigma_{\alpha\gamma}$. Moreover, changing the local sections by a
coboundary is just a reparametrization and yields the same genus-one
fibration after gluing. Therefore, the distinct genus-one fibrations
that can be constructed by twisting the elliptic fibration are in
one-to-one correspondence with the elements of the cohomology group
\begin{equation}
  \sha_B(\mathcal{A}) = H^1(B, \mathcal{A}),
\end{equation}
also known as the Tate-Shafarevich group. By the above discussion, we
can identify its elements with the set of genus-one fibrations having
the same Jacobian, that is, having the same $\tau$.

This construction works as stated whenever the fibration $Y\to B$ is 
generic, i.e., has only Kodaira fibers of types $I_0$, $I_1$ and $II$ (and
no non-Abelian gauge symmetry).  For more complicated fibrations, the
sheaf $\mathcal{A}$ misses too much of the structure of $Y$, and
the more detailed analysis of \cite{MR1242006} must be used.

\subsection{Relation with Discrete Torsion}

On an elliptic curve, that is, a curve of genus one with marked point
$\sigma$, a choice of point $p$ amounts to a choice of line bundle
$\Osheaf(p-\sigma)$ with vanishing first Chern class. Hence, we could
use the defining data of the Tate-Shafarevich group just as well to
glue something \emph{by tensoring with a line bundle} of vanishing
first Chern class. Except for the ``vanishing $c_1$'' part, one
recognizes this as the familiar gerbe data defining the twist of a
projective vector bundle. For torsion gerbe characteristic classes,
this is also knows as discrete torsion in string theory. In more fancy
language, we can think of the sheaf $\mathcal{A}$ as the degree-zero
part of the relative Picard sheaf
\begin{equation}
  0 \longrightarrow
  \mathcal{A} \longrightarrow
  \underline{Pic}(Y/B) \longrightarrow
  \Z \longrightarrow
  0,
\end{equation}
except that the Picard sheaf may not be well-defined for sufficiently
complicated fibrations, leading to considerable technical difficulties
for which we refer the reader to~\cite{MR1242006, MR1918053,
  MR2399730}. Nevertheless, the induced map in cohomology
\begin{equation}
 \sha_B(\mathcal{A}) = H^1(B, \mathcal{A})
  \longrightarrow
  H^1\big(B, \underline{Pic}(Y/B)\big)
\end{equation}
and the Leray spectral sequence 
\begin{equation}
  \label{eq:Leray}
  \cdots \longrightarrow
  Br'(B) \longrightarrow
  Br'(Y) \longrightarrow
  H^1\big(B, \underline{Pic}(Y/B)\big)
  \longrightarrow \cdots
\end{equation} 
for $\Osheaf^\times_Y$ on the projection $Y\to B$ links the
Tate-Shafarevich group to the (cohomological) Brauer group and gerbes
(for generic fibrations).

However, note that only the Brauer group on the total space
\emph{modulo} the pull-back of the Brauer group on the base has a
chance of contributing according to eq.~\eqref{eq:Leray}. That is, the
Tate-Shafarevich group provides similar but strictly independent
global information from B-fields and gerbes in Type \IIB. In the
special case where $\dim B=1$ or where all fibers are irreducible
(hence no non-Abelian gauge symmetry), the relationship between the
Brauer and the Tate-Shafarevich groups becomes particularly simple. In
this case~\cite{MR1242006} the quotient of the Brauer groups is indeed
the only contribution,
\begin{equation}
  \sha_B(\mathcal{A}) = \coker\big( Br'(B) \to Br'(Y) \big).
\end{equation}
Hence, under these simplifying assumptions the \IIB\ gerbes on the
base and the Tate-Shafarevich group of the fibration combine together
into the Brauer group of the total space of the fibration.

The \IIB\ fluxes should, more precisely, be thought of as classes in a
suitable version of K-theory. For example, for orientifolds the
correct flavor of K-theory is KR-theory, Real equivariant with respect
to the orientifold involution. It would be nice to understand this
better and have a direct connection to the Tate-Shafarevich group that
does not proceed via cohomology.

%%% Local Variables:
%%% TeX-master: "Main.tex"
%%% eval: (TeX-PDF-mode 1)
%%% End:

\section{Moduli of Genus-One Fibrations}
\label{sec:hyper_count}

\subsection{Weierstrass Hypersurface}
\label{sec:weierstrass}

So far we have argued that the degrees of freedom in F-theory
include the Tate-Shafarevich group, elements of which correspond
geometrically to distinct genus-one fibrations with the same
Jacobian fibration. Allowing genus-one fibrations
 has direct physical consequences. In this
section, we will see that it corrects the uncharged hypermultiplet
count. In the next section, we will find new ways to break gauge
symmetry. However, before getting ahead of ourselves, let us start with
proper elliptic fibrations giving rise to $SU(5)$ gauge theory to make
contact with physics literature. Only later, starting with
\autoref{sec:no_section}, will we consider genus-one fibrations that
do not admit a section. In order to make use of the strong anomaly
cancellation conditions, let us focus on the case of genus-one fibered
Calabi-Yau threefolds compactifying F-theory down to $6$ dimensions.

The simplest way to construct an elliptic fibration over a fixed base
$B$ is as a Weierstrass hypersurface. That is, consider the
$\IP^2[2,3,1]$ bundle
\begin{equation}
  \label{eq:P231bundle}
  X = 
  \IP\Big( 
  \Osheaf_B(-2K) \oplus \Osheaf_B(-3K) \oplus \Osheaf_B 
  \Big)[2,3,1]
  .
\end{equation}
For suitable bases $B$, an anticanonical hypersurface $Y\subset X$ is a
Calabi-Yau threefold with only canonical singularities for which there
exists a smooth Calabi-Yau resolution $\widetilde{Y}\to Y$.
In order to avoid tensionless strings in our models,
we are only interested hypersurfaces whose
resolution has a fibration is flat, i.e., has  only one-dimensional fibers.
If the base is a toric variety with fan
$\Sigma_B$, then the ambient space $X=X_\Sigma$ is also a toric
variety. To construct the polytope for $X$, we note that the anticanonical
hypersurface equation takes the long Weierstrass form
\begin{equation}
  y^2 = x^3  
  + a_1 x y z + a_2 x^2 z^2 + a_3 y z^3 + a_4 x z^4 + a_6 z^6
\end{equation}
with coefficients $a_i \in \Gamma( \Osheaf_B(K^{-i}))$. The hypersurface
equation defines its Newton polytope and we define $\Sigma$ as its
normal fan, that is, the face fan of the dual
polytope.\footnote{Depending on the base $B$, the dual need not be a
  lattice polytope. That is, the Newton polytope need not be
  reflexive. This means that the singular variety does not admit a
  resolution to a smooth Calabi-Yau hypersurface.} For example, if
$B=\IP^2$ then one obtains the weighted projective space
$X=\IP^4[1,1,1,6,9]$ as ambient fourfold. The defining polytope is:
\begin{equation}
  \begin{array}{|c|c|c|ccc|p{0mm}ccc|}
    \hline
    u & v & w & x & y & z & & & & 
    \\ \hline
    9  & 0 & 0 &  0 & -1 & 3 && 2 & 1 & 1 \\
    6  & 0 & 0 & -1 &  0 & 2 && 1 & 1 & 0 \\
    -1 & 1 & 0 &  0 &  0 & 0 && 0 & 0 & 0 \\
    -1 & 0 & 1 &  0 &  0 & 0 && 0 & 0 & 0 \\
    \hline
    \multicolumn{3}{|c|}{\text{three tops}} & 
    \multicolumn{3}{c|}{\text{fiber}} & 
    \multicolumn{4}{c|}{
      \begin{array}{c}\text{facet}\\ \text{interior}\end{array}
      }
    \\
    \hline
  \end{array}
\end{equation}
The toric variety $X$ corresponding to eq.~\eqref{eq:P231bundle}
inherits singularities from the fiber $\IP^2[2,3,1]$. However, for
$B=\IP^2$ there is a unique toric resolution preserving the fibration,
so we will not dwell on this issue.

The choice of ambient space induces additional structure on the
hypersurface beyond that of a generic torus-fibered Calabi-Yau
threefold over $B$. In particular,
\begin{itemize}
\item The toric divisor $V(z) \cap Y$ is a section of $Y$, namely
  \begin{equation}
    s: B\to Y,\quad
    b \mapsto \big(b, [1:1:0]\big).
  \end{equation}
\item The self-intersection of the canonical class on the toric fiber
  $\IP^2[2,3,1]$ equals $6$. Therefore, $-K_X \cap Y$ is a family of
  $6$-sections. It can be chosen to contain  a section as an
  irreducible component, which we can subtract to leave us with a
  $5$-section. The particular choice
  \begin{equation}
    \{y^2 = x^3\} \cap Y
    =
    \big\{ 
    z 
    \big(a_1 x y + a_2 x^2 z + a_3 y z^2 + a_4 x z^3 + a_6 z^5\big)
    = 0
    \big\}
    \cap Y
  \end{equation}
  of $5$-section in this family is called ``the Tate divisor'' in the
  physics literature~\cite{Marsano:2011nn}.
\end{itemize}
A generic Calabi-Yau hypersurface in the projective space bundle $X$
has a nonsingular total space and
only three types of Kodaira fibers: $I_0$ (a nonsingular fiber),
$I_1$ (a fiber with a node), and $II$ (a fiber with a cusp). 
In particular, its F-theory
compactification has no gauge symmetry at all, only $h^{2,1}(Y)+1=273$
uncharged hypermultiplets. In order to generate non-Abelian
interactions, one needs to find special loci in the complex structure
moduli space where more complicated Kodaira fibers appear. At the same
time, one has to ensure that there still exists a flat
resolution. Examples of terminal singularities as well as examples of
canonical singularities whose resolution is not flat exist. While not
insurmountable obstacles, these are very real complications.

\subsection{Elliptic Fibration with a Toric SU(5)}
\label{sec:toricSU5}

Instead of trying to specialize the Weierstrass equation by hand, we
can also specialize the ambient space in a way that enforces
particular Kodaira fibers. In particular, this avoids potential
terminal singularities: A \emph{generic} three-dimensional Calabi-Yau
hypersurface in a toric variety can always be resolved into a smooth
threefold~\cite{1993alg.geom.10003B}.

For example, consider the following split $I_5$ toric elliptic
fibration over $\IP^2$~\cite{Braun:2011ux, MR1463052, Grassi:2012qw}
with Hodge numbers $h^{1,1}=6$, $h^{2,1}=171$. The ambient space is the
toric variety with polytope $\nabla$ whose points are
\begin{equation}
  \label{eq:nabla_SU5}
  \begin{array}{|c|c|ccccc|ccc|ccc|}
    \hline
    u & v & w_0 & w_1 & w_2 & w_3 & w_4 &
    f_{0}&f_{1}&f_{2} &&&
    \\ \hline
    6 & 0 & 3 & 2 & 1 & 0 & 1 & 3 & 0 & -1 & 1 & 1 & 2 \\
    4 & 0 & 2 & 1 & 0 & 0 & 1 & 2 & -1 & 0 & 0 & 1 & 1 \\
    -1 & 0 & 1 & 1 & 1 & 1 & 1 & 0 & 0 & 0 & 0 & 0 & 0 \\
    -1 & 1 & 0 & 0 & 0 & 0 & 0 & 0 & 0 & 0 & 0 & 0 & 0 \\
    \hline
    \multicolumn{7}{|c|}{\text{three tops}} & 
    \multicolumn{3}{c|}{\text{fiber}} & 
    \multicolumn{3}{c|}{\text{facet interior}}
    \\
    \hline
  \end{array}
\end{equation}
and which is fibered over $\IP^2$ via the projection to the last two
coordinates. In terms of homogeneous coordinates, this is the map
\begin{figure}[htbp]
  \centering
  \raisebox{-0.5\height}{
    \includegraphics{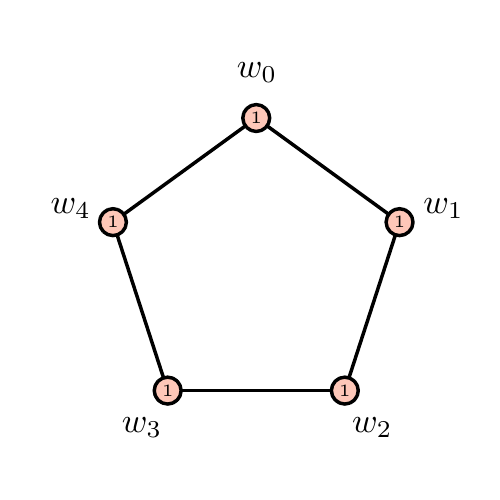}}%
  \raisebox{-0.5\height}{
    \includegraphics[width=0.7\textwidth]{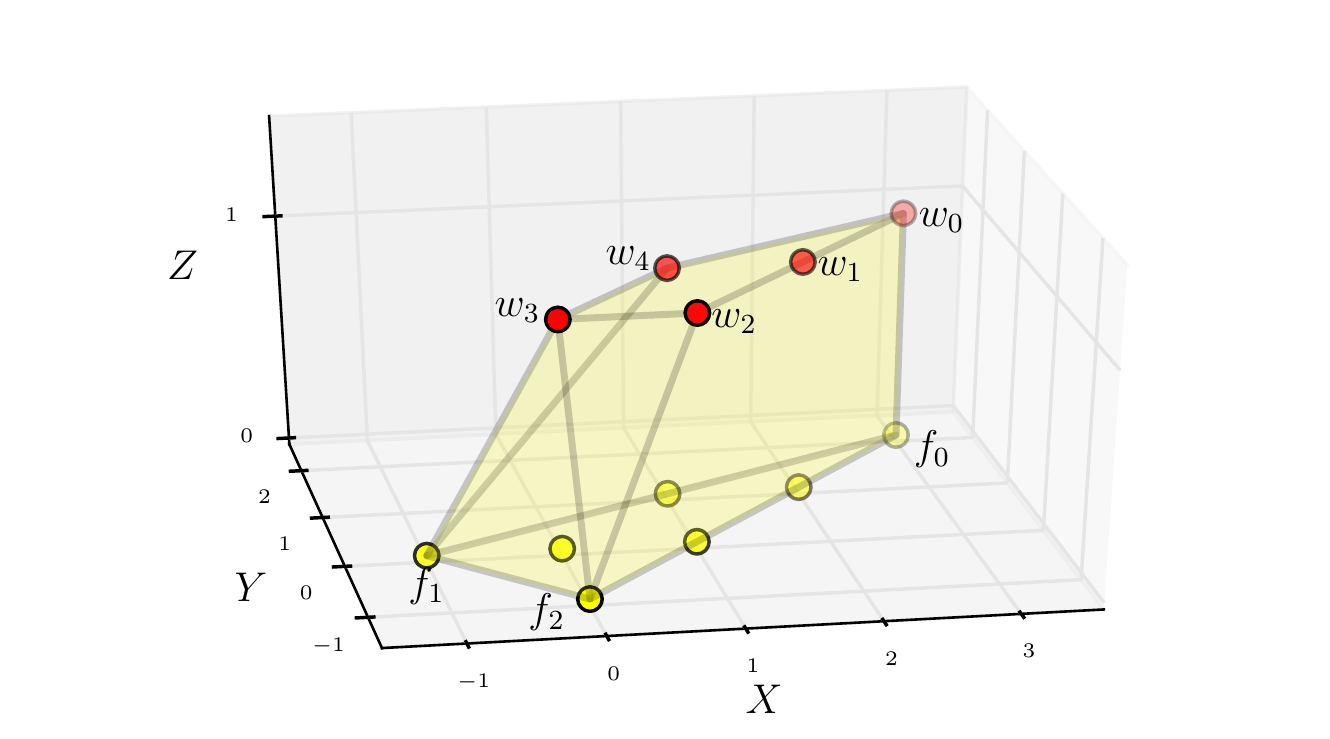}}
  \caption[Split $SU(5)$ top and associated $I_5$ Kodaira fiber.]
  {The split $SU(5)$ top and associated extended Dynkin
    diagram, the $I_5$ Kodaira fiber.}
  \label{fig:top_SU5}
\end{figure}
\begin{equation}
  \pi: X_{\nabla} \to \IP^2
  ,\quad
  [u:v:w_0:\cdots:w_4:f_0:f_1:f_2] \mapsto
  \big[u:v:w_0 w_1 w_2 w_3 w_4 \big]
\end{equation}
The Dynkin diagram of the gauge group can be seen from the
\emph{tops}~\cite{Candelas:1997pq, Candelas:1997jz, Candelas:1997eh,
  Candelas:1997pv, Candelas:2012uu, Braun:2013yti, Braun:2013nqa} of
the fibration. We recall their definition and important properties:
\begin{itemize}
\item Given a ray $\rho$ of the base fan, the preimage $\pi^{-1}(\rho)
  \subset \nabla$ is called a ``top''.
\item Regardless of the dimension of the codimension-two fibered toric
  variety, a top is always a 3-dimensional lattice polytope with the
  origin on one facet (namely the fiber polygon).
\item The top over $\rho$ defines the gauge group over the
  discriminant component $V(\rho)$ as follows:
  \begin{itemize}
  \item The edge graph (with corresponding integral points) not meeting
    the fiber polygon equals the quotient of the Dynkin diagram by the
    monodromy.
  \item The Dynkin labels are the height of the points over the fiber
    polygon.
  \end{itemize}
\item The toric (multi-)sections correspond to the vertices of the
  fiber polygon, and their intersection with the fiber irreducible
  components are also visible as the edges of the top.
\end{itemize}
In the case of an elliptically fibered K3, there are only two rays in
the base $\IP^1$ and they divide the K3 polytope in a top and a bottom
half. This is the origin of the name, but in general there is one top
for each ray in the base.

In our example there are three tops over the three rays of the fan of
the base $\IP^2$. Two of them are trivial with only a single vertex
added over the fiber polygon. The third accounts for the $SU(5)$ gauge
group and is shown in \autoref{fig:top_SU5}. Using toric geometry to
translate the polyhedral data into an algebraic
variety~\cite{2000math.....10082H, Braun:2011ux,
  BraunNovoseltsev:toric_variety}, each integral point of the $w$-top
corresponds to an irreducible component of the (complex
two-dimensional) toric fiber over $w=0$ in the base. The hypersurface
equation cuts out $\IP^1$'s in each irreducible toric fiber
component. In the case at hand, the hypersurface equation yields a
single $\IP^1$ in each of the five components, linked as in the $I_5$
Kodaira fiber.

\subsection{Adding a single U(1)}
\label{sec:single_U1}

For physics applications it is desirable to have a non-trivial
Mordell-Weil group, unlike the threefold constructed in
\autoref{sec:toricSU5}. The torsion part of the Mordell-Weil group is
a discrete symmetry of the low energy effective action, and the rank
equals the number of $U(1)$ factors in the gauge group. Both have
important phenomenological applications, for example ruling out
certain operators that would lead to excessive proton decay. Although
not necessary for the remainder of this paper, we now make a small
digression to discuss ways to realize additional $U(1)$ factors. The
reader not interested in these can skip ahead to
\autoref{sec:no_section}.
\begin{sidewaystable}[htbp]
  \centering
  \begin{tabular}{cc}
    \begin{math}
      \displaystyle
      \begin{array}{|c|c|ccccc|cccc|ccc|}
        \hline
        u & v & w_0 & w_1 & w_2 & w_3 & w_4 &
        f_{0}&f_{1}&f_{2} & f_{3} &&&
        \\ \hline
        6 & 0 & 3 & 2 & 1 & 0 & 1 & 3 & 0 & -1 & -1 & 2 & 1 & 1 \\
        4 & 0 & 2 & 1 & 0 & 0 & 1 & 2 & -1 & -1 & 0 & 1 & 1 & 0 \\
        -1 & 0 & 1 & 1 & 1 & 1 & 1 & 0 & 0 & 0 & 0 & 0 & 0 & 0 \\
        -1 & 1 & 0 & 0 & 0 & 0 & 0 & 0 & 0 & 0 & 0 & 0 & 0 & 0 \\
        \hline
        \multicolumn{7}{|c|}{\text{three tops}} & 
        \multicolumn{4}{c|}{\text{fiber}} & 
        \multicolumn{3}{c|}{\text{facet int.}}
        \\
        \hline
      \end{array}
    \end{math}
    &
    \hspace{-1cm}
    \raisebox{-0.5\height}{
      \includegraphics{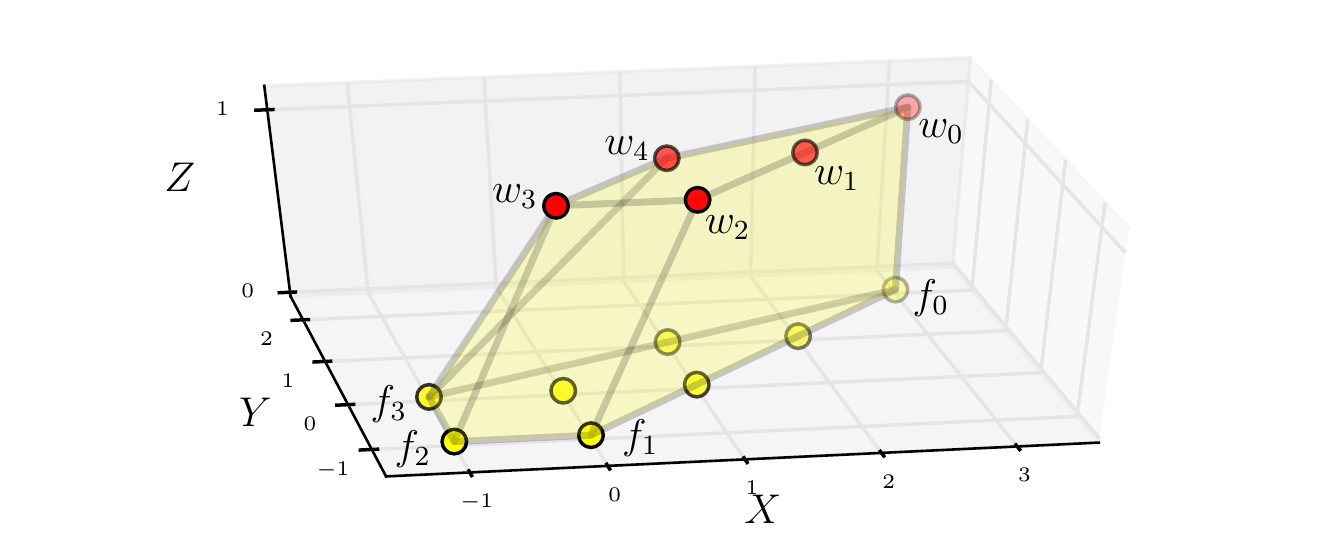} 
    }
    \\
    \begin{math}
      \displaystyle
      \begin{array}{|c|c|ccccc|cccc|cc|}
        \hline
        u & v & w_0 & w_1 & w_2 & w_3 & w_4 &
        f_{0}&f_{1}&f_{2} & f_{3} &&
        \\ \hline
        6 & 0 & 3 & 2 & 1 & 0 & 1 & 3 & 1 & -1 & -1 & 1 & 2 \\
        4 & 0 & 2 & 1 & 0 & 0 & 1 & 2 & 0 & -1 & 0 & 1 & 1 \\
        -1 & 0 & 1 & 1 & 1 & 1 & 1 & 0 & 0 & 0 & 0 & 0 & 0 \\
        -1 & 1 & 0 & 0 & 0 & 0 & 0 & 0 & 0 & 0 & 0 & 0 & 0 \\ 
        \hline
        \multicolumn{7}{|c|}{\text{three tops}} & 
        \multicolumn{4}{c|}{\text{fiber}} & 
        \multicolumn{2}{c|}{\text{facet int.}}
        \\
        \hline
      \end{array}
    \end{math}
    & 
    \hspace{-1cm}
    \raisebox{-0.5\height}{
      \includegraphics{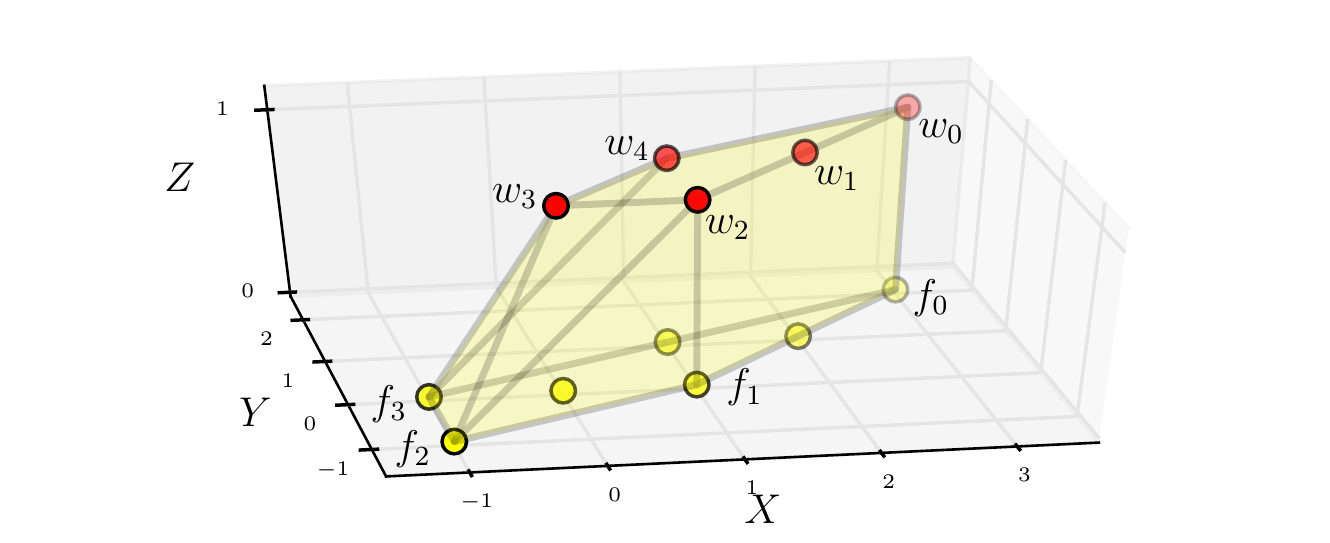}
    }
    \\
    \begin{math}
      \displaystyle
      \begin{array}{|c|c|ccccc|cccc|c|}
        \hline
        u & v & w_0 & w_1 & w_2 & w_3 & w_4 &
        f_{0}&f_{1}&f_{2} & f_{3} &
        \\ \hline
        6 & 0 & 3 & 2 & 1 & 0 & 1 & 3 & 2 & -1 & -1 & 1 \\
        4 & 0 & 2 & 1 & 0 & 0 & 1 & 2 & 1 & -1 & 0 & 1 \\
        -1 & 0 & 1 & 1 & 1 & 1 & 1 & 0 & 0 & 0 & 0 & 0 \\
        -1 & 1 & 0 & 0 & 0 & 0 & 0 & 0 & 0 & 0 & 0 & 0 \\
        \hline
        \multicolumn{7}{|c|}{\text{three tops}} & 
        \multicolumn{4}{c|}{\text{fiber}} & 
        \multicolumn{1}{c|}{\text{facet int.}}
        \\
        \hline
      \end{array}
    \end{math}
    &
    \hspace{-1cm}
    \raisebox{-0.5\height}{
      \includegraphics{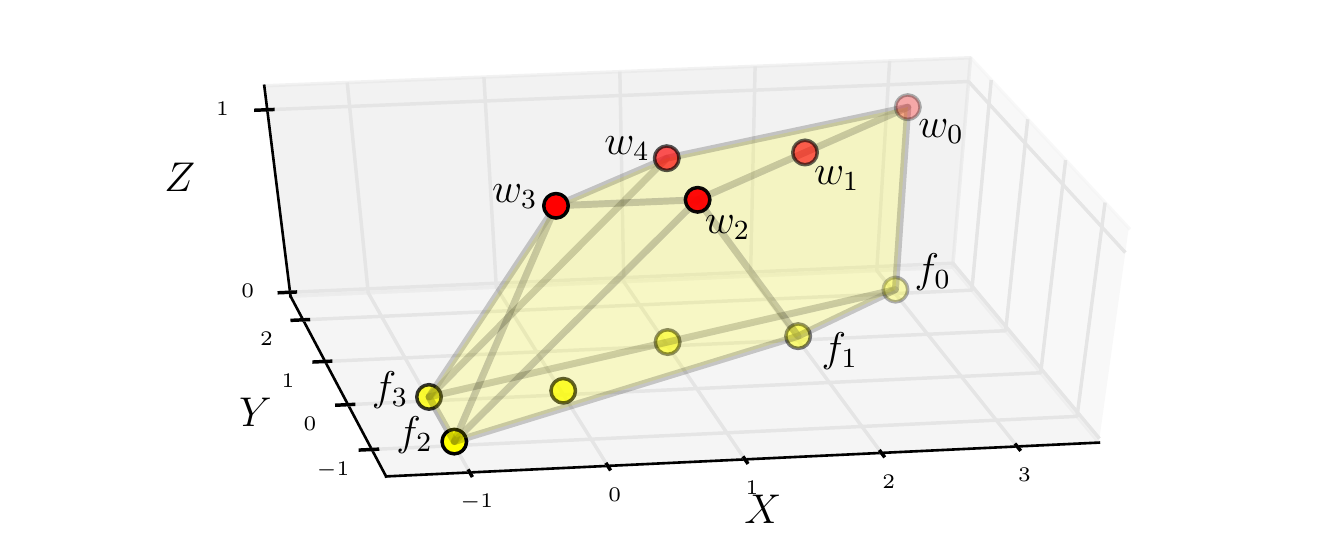}
    }
  \end{tabular}
  \caption[The three $SU(5) \times U(1)$ models with $h^{1,1}(Y)=7$,
  $h^{2,1}(Y)=109$.]
  {The three $SU(5) \times U(1)$ models with $h^{1,1}(Y)=7$,
    $h^{2,1}(Y)=109$ and the corresponding $SU(5)$ top.}
  \label{tab:SU5xU1}
\end{sidewaystable}

The most straightforward idea to generate a $U(1)$ is to take the
Weierstrass hypersurface and restrict the complex structure further
until an extra section appears. (In fact, as argued in \cite{MW},
this is the only way to do it.)  If one wants the additional $U(1)$ to
be toric, then this means one should find a point to add to the fiber
polytope such that it induces a new toric section and such that the
convex hull is again reflexive. It is easy to see that adding
$(-1,-1,0,0)$ to the polytope in eq.~\eqref{eq:nabla_SU5} works, and
the new polytope is drawn in the top row of \autoref{tab:SU5xU1}. In
fact, the new Calabi-Yau hypersurface has three sections: $V(f_0)$,
$V(f_1)$, and $V(f_2)$. The new $h^{1,1}(Y)=7$, so the additional toric
sections generate a rank-one Mordell-Weil group. Therefore, the gauge
group is enhanced to $SU(5)\times U(1)$. The self-intersection number
of the canonical class of the generic toric fiber drops from $6$ on
$\IP^2[2,3,1]$ to $5$. Hence the Tate divisor degenerated into a
$4$-section and the new section. This construction is known as the
``U(1)-restricted Tate model'' in the physics literature.

This is not the only way to generate a single $U(1)$ using toric 
techniques,\footnote{A systematic analysis of the toric ambient spaces
  for elliptic curves which can be employed to construct multiple
  toric sections for elliptically-fibered varieties was given in
  \cite{Grassi:2012qw, Cvetic:2013nia, Cvetic:2013uta, Cvetic:2013jta,
    Cvetic:2013qsa}.} and in
\autoref{tab:SU5xU1} we list two more polytopes that give rise to the
same spectrum. In fact, it is easy to see that the top two cases
contain the bottom polytope, that is, the hypersurfaces of the top two
toric varieties are special limits of the hypersurface of the bottom
toric variety. In particular, we recognize that the extra $U(1)$ comes
from the additional vertex of the fiber polytope. This extra vertex
yields an additional integral point of the $\nabla$-polytope not in
the interior of a facet, and therefore increases $h^{1,1}$ by
one. Hence, the reason for the additional $U(1)$ is this additional
vertex in the toric picture, and not the fact that one of the three is
the restriction of the Weierstrass hypersurface to a special
point. That the three polytopes are contained in each other implies
that the Weierstrass model is the same, but the details of the
resolution to a smooth threefold are different. In particular, the
toric description of the codimension-two degeneration where the
$\Rep{5}_3$ matter is localized differs. 
\begin{itemize}
\item In the first $SU(5)\times U(1)$ polytope of \autoref{tab:SU5xU1}
  we recognize a conifold singularity defined by the square $f_1$,
  $f_2$, $w_3$, $w_2$. In constructing a smooth Calabi-Yau threefold,
  one must resolve it in one of two ways. Either way, this extra curve
  is the additional $\CP^1$ in $7$ out of the $19$ fibers where the
  $I_5$ Kodaira fiber degenerates into an $I_6$ Kodaira
  fiber. Depending on which resolution one takes, this curve is
  contained in the $V(f_1)\cap Y$ or the $V(f_2)\cap Y$ section.
\item In the second and third $SU(5)\times U(1)$ polytope of
  \autoref{tab:SU5xU1}, none of the extra $\CP^1$ in the
  codimension-two $I_6$ fibers are toric curves in the top. All toric
  sections intersect the irreducible components of $I_6$
  codimension-two fibers transversely.
\end{itemize}
None of these differences in resolution impact the physics of the
$SU(5)\times U(1)$ F-theory model.

\subsection{U(1) Charge Assignments}
\label{sec:U1charges}

In the case at hand all charged hypermultiplets arise from isolated
curves. Their $U(1)$ charge of the hypermultiplet on a curve $C$ under
the $U(1)$ of a section $S\in MW(Y)$ is determined by the intersection
\begin{equation}
  C \cdot \pi(S) 
  = 
  C\cdot S
  + \sum_{i,j}
  (S\cdot \theta_i) (A^{-1})_{ij} (\Theta_j\cdot C)
\end{equation}
where the $\theta_i$ are the curves carrying the roots, that is, the
irreducible components of the gauge group discriminant component not
intersecting the given section, the $\Theta_j$ are the divisors swept
out by the curve $\theta_i$ over the discriminant, and $A$ is the
Cartan matrix of the gauge group.
\begin{figure}[htbp]
  \centering
  \includegraphics{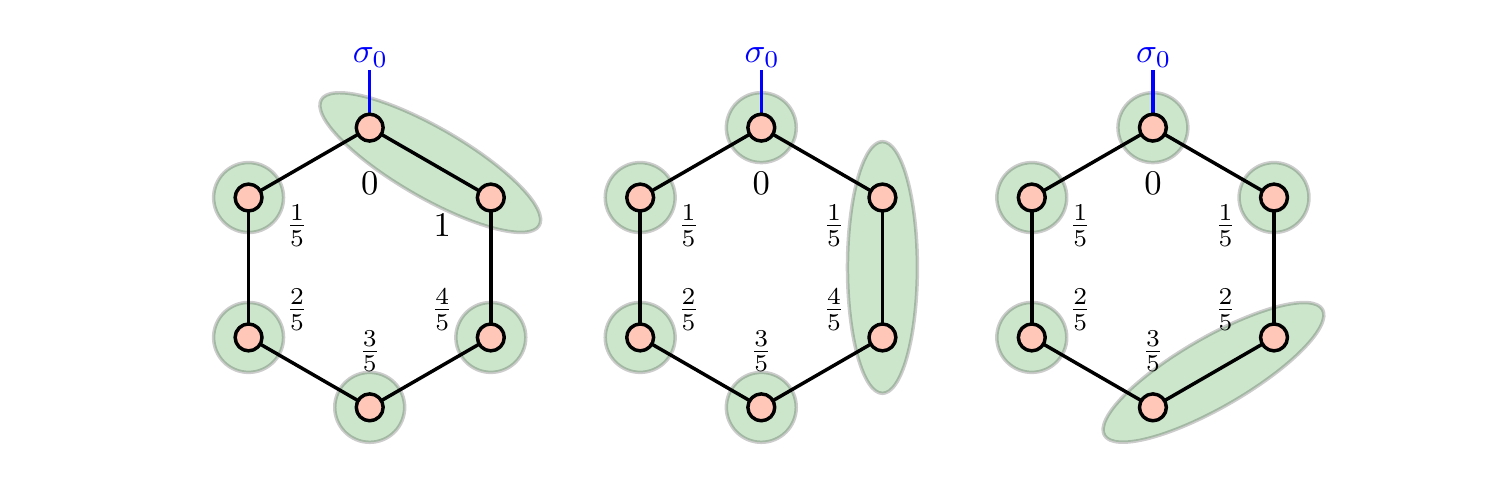}
  \vspace{-1cm}
  %%% the \Rep macro is broken in a figure caption
  \caption[$U(1)$ charges for $SU(5)\times U(1)$ fundamental
  hypermultiplets.]
  {The $U(1)$ charge of a $\Rep{5}$ hypermultiplet localized at a
    codimension-two fiber where the $I_5$ discriminant (green)
    degenerates to $I_6$ (red). Pick the red node where the section
    generating the Mordell-Weil group intersects the fiber. The number
    next to the node is the ratio of the $U(1)$ charge relative to a
    $\Rep{1}$ hypermultiplet.}
  \label{fig:SU5U1charge5}
\end{figure}
For example, for the $\Rep{5}$ hypermultiplets the $U(1)$ charge works
out as in \autoref{fig:SU5U1charge5}.\footnote{Note that we will
  always normalize $U(1)$ charges to be integral. The electron charge
  is $-3$.} 

In addition to the $SU(5)$-charged hypermultiplets, there are also a
number of hypermultiplets charged only under the $U(1)$ whose curves
are localized away from the non-Abelian discriminant component. That
is, they are localized at special points of the $I_1$ discriminant
component. We will only encounter the three simplest cases in the
following discussion:
\begin{itemize}
\item Smooth point: $u=v$ or $u=v^2$ \hfill (Milnor number 0)
\item Node: $uv=0$ \hfill (Milnor number 1)
\item Cusp: $u^2 = v^3$ \hfill (Milnor number 2)
\end{itemize}
It is of practical importance to be able to determine the number of
such singularities in the $I_1$ component efficiently for a generic
hypersurface. One useful trick is to take the polynomial
$\delta_1(u,v,w)$ whose vanishing defines $I_1$ and compute its
discriminant with respect to one of the variables. This then
factorizes into linear, quadratic, and cubic factors that can be
counted easily. For example, the $SU(5)$ model has $171$ cusps in the
$I_1$ discriminant component and no nodes. In fact, without any
section there cannot be any node for intersection-theoretic reasons:
The resolved Calabi-Yau would have an $I_2$ Kodaira fiber over the
node, but without a section there is no divisor available to be the
Poincar\'e dual of the irreducible fiber component not intersecting
the given section.

The three $SU(5)\times U(1)$ models with $h^{1,1}=7$, $h^{2,1}=109$, see
\autoref{tab:SU5xU1}, have $171$ cusps and $63$ nodes in the $I_1$
discriminant component. Computing the intersections with the toric
sections, one finds the $U(1)$ charges of the hypermultiplets to be
\begin{equation}
  7 \times \Rep{5}_3 
  ~\oplus~
  12 \times \Rep{5}_2
  ~\oplus~
  3 \times \Rep{10}_1
  ~\oplus~
  63 \times \Rep{1}_5
\end{equation}
which satisfies the $U(1)$ anomaly cancellation conditions for
an appropriate choice of Green-Schwarz term.
For completeness, let us recall the anomaly
cancellation conditions for a single non-Abelian gauge group $G$ and a
single $U(1)$ when the base $B$ is $\mathbb{P}^2$. A Green-Schwarz term
proportional to
\begin{equation}
\int -\frac32 \tr R^2 + 2b \tr_G F_G^2 + 2 \tilde b F_{U(1)}^2
\end{equation}
will cancel the anomalies provided that~\cite{Park:2011wv, MW}:
\begin{equation}
  \begin{gathered}
  \begin{aligned}
    18 b =& \sum_i A_{R_i} - A_{\Ad}, &    
    0 =& \sum_i B_{R_i} - B_{\Ad}, &
    3 b^2 =& \sum_i C_{R_i} - C_{\Ad},
  \end{aligned}
  \\
  \begin{aligned}
    0 =& \sum_i E_{R_i} r_i, &\qquad
    18 \tilde{b} =& \sum_i \dim(R_i) r_i^2, \\
    b \tilde{b} =& \sum_i A_{R_i} r_i^2, &
    3 \tilde{b}^2 =& \sum_i \dim(R_i) r_i^4.
  \end{aligned}
  \end{gathered}
\end{equation}
where the model has hypermultiplets transforming in the
representations $(R_i,r_i)$.
In the example at hand, all anomalies cancel\footnote{The non-Abelian
  anomaly coefficients, see \autoref{sec:sage}, are
  $A_\Rep{1}=B_\Rep{1}=C_\Rep{1}=0$, $A_\Rep{5}=B_\Rep{5}=1$,
  $C_\Rep{5}=0$, $A_\Rep{10}=-B_\Rep{10}=C_\Rep{10}=3$, and
  $A_{\Ad}=B_{\Ad}=10$, $C_{\Ad}=6$.} for $b=1$ and $\tilde{b}=120$.

\subsection{Three U(1)'s}
\label{sec:SU5xU1cubed}

In order to realize three toric $U(1)$ factors in a toric
hypersurface, the generic ambient space fiber must be
$\dP6$~\cite{Grassi:2012qw, Braun:2013nqa, Cvetic:2013qsa}.  One
possibility for $SU(5)\times U(1)^3$ gauge symmetry with $h^{1,1}=9$,
$h^{2,1}=52$ is:
\begin{equation}
  \label{eq:toricSU5U1cubed}
  \displaystyle
  \begin{array}{|c|c|ccccc|cccccc|}
    \hline
    u & v & w_0 & w_1 & w_2 & w_3 & w_4 &
    f_{0}&f_{1}&f_{2} & f_{3} & f_4 & f_5
    \\ \hline
    3 & 0 & 0 & -1 & -1 & 0 & 0 & 1 & 1 & 0 & -1 & -1 & 0 \\
    1 & 0 & -1 & -1 & 0 & 1 & 0 & 1 & 0 & -1 & -1 & 0 & 1 \\
    -1 & 0 & 1 & 1 & 1 & 1 & 1 & 0 & 0 & 0 & 0 & 0 & 0 \\
    -1 & 1 & 0 & 0 & 0 & 0 & 0 & 0 & 0 & 0 & 0 & 0 & 0 \\
    \hline
    \multicolumn{7}{|c|}{\text{three tops}} & 
    \multicolumn{6}{c|}{\text{fiber}} 
    \\
    \hline
  \end{array}
\end{equation}
\begin{figure}[h]
  \centering
  \includegraphics{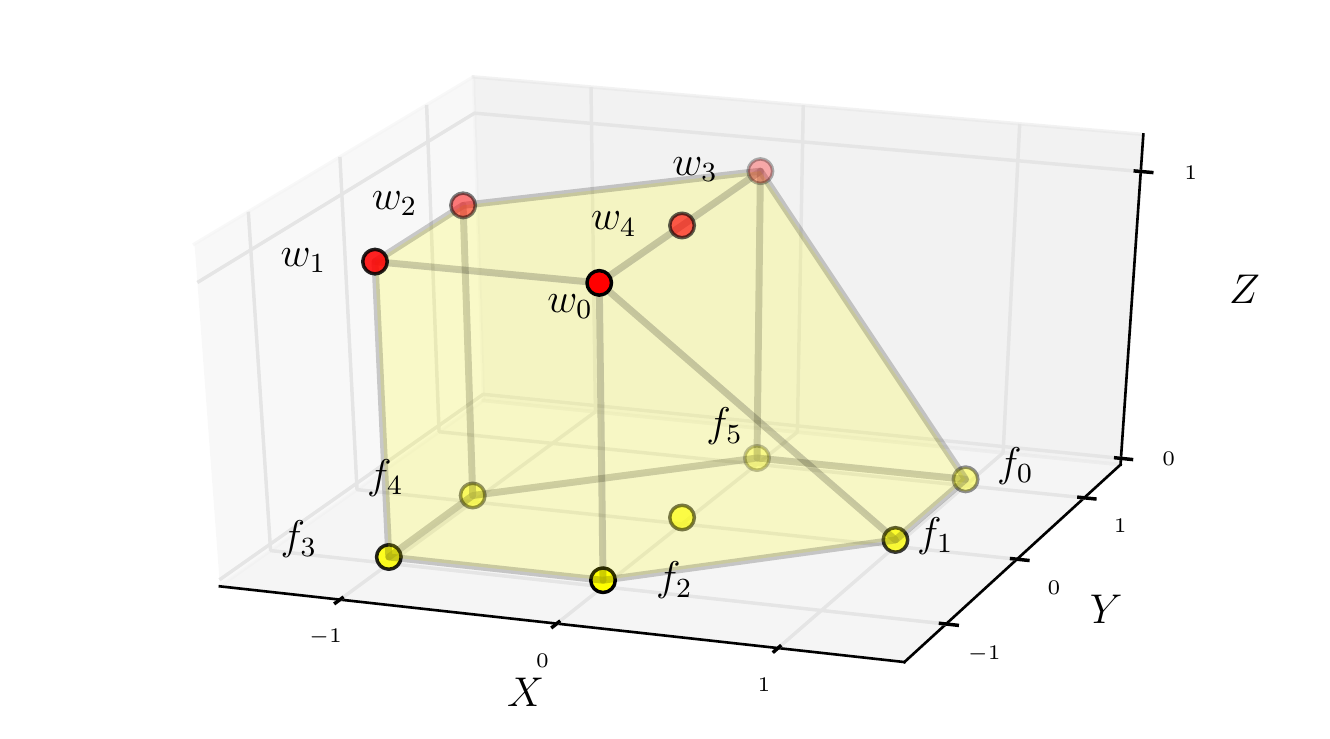}
  \caption{The $SU(5)\times U(1)^3$ top.}
  \label{fig:top_SU5xU1cubed}
\end{figure}%
All $SU(5)$ models studied so far have $19$ points on the discriminant
where the vanishing degree jumps from $(0,0,5)$ to $(0,0,6)$ and three
points where it jumps to $(2,3,7)$. Therefore, the $SU(5)$-charged
hypermultiplets are $19\times \Rep{5} \oplus 3 \times \Rep{10}$ which
is precisely what is needed to cancel the non-Abelian gauge anomaly
\begin{equation}
  18 b = 19\cdot 1 + 3\cdot 3 - 10
  ,\quad
  0 = 19\cdot 1 + 3\cdot (-3) - 10
  ,\quad
  3 b^2 = 19\cdot 0 + 3\cdot 3 - 6
  .
\end{equation}
The $SU(5)\times U(1)^3$ model with toric sections, see
eq.~\eqref{eq:toricSU5U1cubed}, has $171$ cusps and $122$ nodes.

\subsection{No Section}
\label{sec:no_section}

After the detour on F-theory models with a section, we now finally
present an example \emph{without} a section. In addition to the
$SU(5)$ model with a toric section, see eq.~\eqref{eq:nabla_SU5},
there are $15$ further fibered toric hypersurfaces with an $I_5$
discriminant component and $h^{1,1}=6$, that is, no $U(1)$. They do
differ in the number of complex structure parameters, ranging from
$90$ to $49$ instead of $h^{2,1}=171$ of eq.~\eqref{eq:nabla_SU5}. 
\begin{table}
  \centering
  \begin{tabular}{c|ccc}
    $G$ & $h^{1,1}$ & $h^{2,1}_\text{expected}$ & $h^{2,1}$ \\
    \hline
    $SU(2)$ & 3 & 231 & 
    $\{231,~123,~119,~111^2,~107,~81,$ \\
    &&& $\quad 77,~76,~75,~73,~72,~71\}$ \\
    $SU(3)$ & 4 &  208 & 
    $\{208,~110,~104,~100,~98,$ \\
    &&& $\quad 73,~71,~68,~65,~64^4,~61 \}$ \\
    $SU(4)$ & 5 &  189 & 
    $\{189,~99,~98,~93,~90,~89,~87^2,~83,$ \\
    &&& $\quad 66,~62,~60,~59^3,~58,~57^2,~56,~55 \}$ \\
    $SU(5)$ & 6 & 171 & 
    $\{171,~90,~87,~84,~83,~81,~80,~78,$ \\
    &&& $\quad 57,~54^2,~53,~52,~51^2,~50,~49 \}$ \\
    $SU(6)$ & 7 &  154 & 
    $\{154,~151,~79^2,~77,~74,~73,~71^3,$ \\
    &&& $\quad 68^2,~64,~47^4,~46^2,~43^2,~40 \}$ \\
    $SU(7)$ & 8 &  138 & 
    $\{138,~68,~66^2,~62,~60,~42,~41,~38 \}$ \\
    $SU(8)$ & 9 &  123 & 
    $\{123,~60,~59,~54,~51^2,~47,~36 \}$ \\
    $SU(9)$ & 10 & 109 & 
    $\{109,~53,~46,~44 \}$ \\
    $SU(10)$ & 11 & 96 & 
    $\{96,~47,~39,~32 \}$ \\
    $SU(11)$ & 12 & 84 & 
    $\{84,~30 \}$ \\
    $SU(12)$ & 13 & 73 & 
    $\{73 \}$ \\
    $SU(13)$ & 14 & 63 & 
    $\{63 \}$ \\
    $SU(14)$ & 15 & 54 & 
    $\{54 \}$ \\
    $SU(15)$ & 16 & 46 & 
    $\{46 \}$ \\
    $SU(16)$ & 17 & 39 & 
    $\{39 \}$ \\
    $SU(17)$ & 18 & 33 & 
    $\{33 \}$ \\
    $SU(24)$ & 25 & 19 & 
    $\{19 \}$ \\
  \end{tabular}
  \caption[Hodge numbers of the toric CY-threefolds over $\CP^2$.]
  {Hodge numbers of the Calabi-Yau threefolds in toric
    varieties, fibered over $\CP^2$, with gauge group a pure $SU(n)$
    such that the non-Abelian discriminant component is a toric curve.
    The expected number of parameters $h^{2,1}_\text{expected}
    = 271 - 23 n + \frac{n(n+1)}{2}$. Exponents indicate that the
    value of $h^{2,1}$ is realized by multiple polytopes. Only the hypersurfaces with
    $h^{2,1}_\text{expected}$ complex structure moduli are actual
    elliptic fibrations, others are genus-one fibrations without section.} 
  \label{tab:single_block_SU5}
\end{table}
This is quite a common phenomenon. In \autoref{tab:single_block_SU5}
we list all toric hypersurfaces fibered over $\CP^2$ such that the
gauge group is only $SU(n)$ and the non-Abelian discriminant component
is a toric curve in the base $\CP^2$. Just to be explicit, we will use
the $SU(5)$ model with $h^{2,1}=90$ in the remainder of this section as
an example. The polytope of the ambient toric variety is
\begin{equation}
  \label{eq:nabla_SU5_nosection}
  \displaystyle
  \begin{array}{|c|c|ccccc|ccc|c|}
    \hline
    u & v & w_0 & w_1 & w_2 & w_3 & w_4 &
    f_{0} & f_{1}&f_{2} &
    \\ \hline
    2 & 0 & 1 & 1 & 0 & -1 & 0 & 1 & 1 & -1 & 1 \\
    1 & 0 & 1 & 0 & 0 & 1 & 1 & 1 & -1 & 0 & 0 \\
    -1 & 0 & 1 & 1 & 1 & 1 & 1 & 0 & 0 & 0 & 0 \\
    -1 & 1 & 0 & 0 & 0 & 0 & 0 & 0 & 0 & 0 & 0 \\
    \hline
    \multicolumn{7}{|c|}{\text{three tops}} & 
    \multicolumn{3}{c|}{\text{fiber}} & 
    \text{facet int.}
    \\
    \hline
  \end{array}
\end{equation}
\begin{figure}[htbp]
  \centering
  \includegraphics{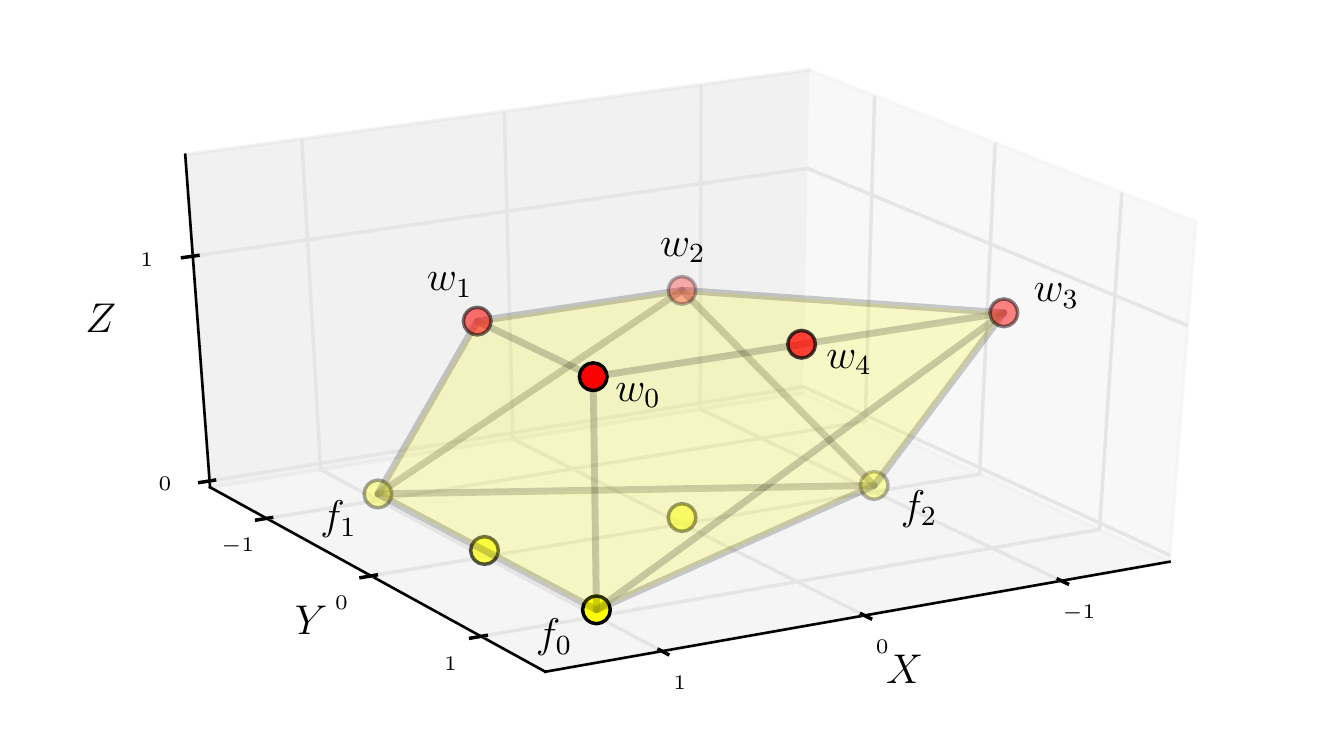}
  \caption{The $SU(5)$ top without section.}
  \label{fig:top_SU5_nosection}
\end{figure}%
with the fibration being defined by the projection on the last two
coordinates. The fiber polygon is that of $\IP^2[1,1,2]$ which does
not lead to a toric section, see \autoref{fig:top_SU5_nosection}. In
fact, there is no section at all: By a direct computation, the $I_1$
discriminant component of a generic hypersurface can be seen to have
$171$ cusps and $81$ nodes. As we mentioned earlier, the presence of a
node means that a single section cannot serve as Poincar\'e dual
divisor to both irreducible components of the $I_2$ resolution. Hence
there cannot be a section. However, a two-section is clearly allowed
since it can easily intersect both components of the $I_2$. And, in
fact, the $\IP^2[1,1,2]$ fiber polytope does induce 2 toric
two-sections, $V(f_0)\cap Y$ and $V(f_1)\cap Y$. Each of them meets
the $I_5$ Kodaira fiber of the non-Abelian discriminant component in
two distinct components.

The Calabi-Yau threefold $Y$ is a genus-one fibration without a
section and therefore not an elliptic fibration. However, it has
a Jacobian fibration $J(Y)$ and $J(Y)$ in turn 
has a Weierstrass model $W_Y = \{y^2 = x^3+f x+g\}$.  (A
Weierstrass model always has a section by construction.) The Weierstrass
model also has $81$ terminal singularities over the $81$ nodes in the
$I_1$ discriminant component, so it is too singular to define a string
theory or M-theory compactification. 
However, as described in \autoref{sec:singularities},
these terminal singularities come along with a discrete torsion
class in the cohomology of the big blowup, and the M-theory
vacuum with $3$-form flux on that class is expected to be a sensible
M-theory model with 81 ``frozen'' singularities.

To further verify that compactifying F-theory
on a genus-one fibration without section makes sense, we now
turn to the anomaly cancellation conditions which provide a very
stringent consistency check.

\section{Gravitational Anomaly Cancellation}
\label{sec:anomaly}

Na\"{\i}vely, the genus-one fibrations are at odds with the gravitational
anomaly cancellation \cite{Berglund:1998rq}.
With fixed gauge group and changed matter
content, the number of uncharged hypermultiplets must satisfy
$H_u+H_c-V+29T = 273$. The standard lore is that the uncharged
hypermultiplets are the $h^{2,1}(Y)$ complex structure moduli plus one
universal hypermultiplet, for a total of $H_u=h^{2,1}(Y)+1$. This
cancels the anomaly for an elliptic fibration as it is related to its
Euler number~\cite{Grassi:2000we,Grassi:2011hq}. A genus-one fibration has fewer
complex structure moduli and, at least naively, not enough uncharged
hypermultiplets. In fact, the missing complex structure moduli are
easily understood from the nodes in the $I_1$ discriminant component:
The complex structure moduli determine the position of the $I_1$, and
each node imposes one additional constraint. For example, the
genus-one fibration from \autoref{sec:no_section} has $h^{2,1}(Y)=90$
and $81$ nodes in the $I_1$. If we add those integers, we arrive at
$171$ which is equal to the number of complex structure moduli of the
$SU(5)$ elliptic fibration eq.~\eqref{eq:nabla_SU5}. This suggests
that the uncharged hypermultiplet count for genus-one fibrations is
corrected to
\begin{equation}
  H_u = h^{2,1}(Y)+1+\#\{\text{nodes}\}
  .
\end{equation}

In fact, Witten's quantization argument for rigid
curves~\cite{Witten:1996qb} tells us that there is an additional
hypermultiplet localized at the $I_2$ fiber over the node, so this
correction is to be expected. What is new is that this localized
hypermultiplet is \emph{uncharged}.
\begin{figure}[htbp]
  \centering
  \begin{minipage}{0.45\linewidth}
    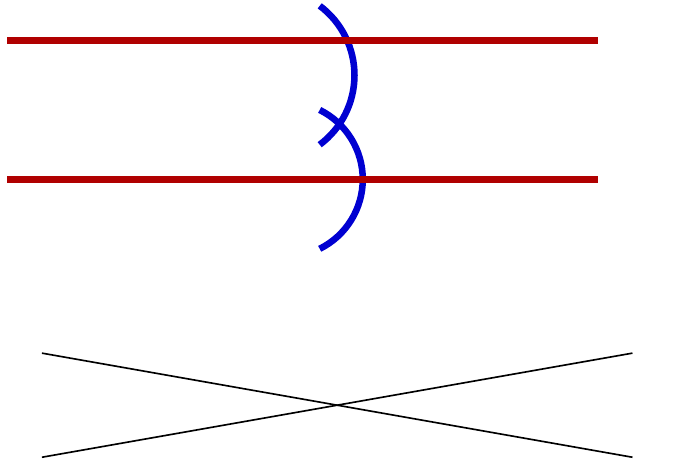
  \end{minipage}
  \hfill
  \raisebox{0.4mm}{
    \begin{minipage}{0.45\linewidth}
      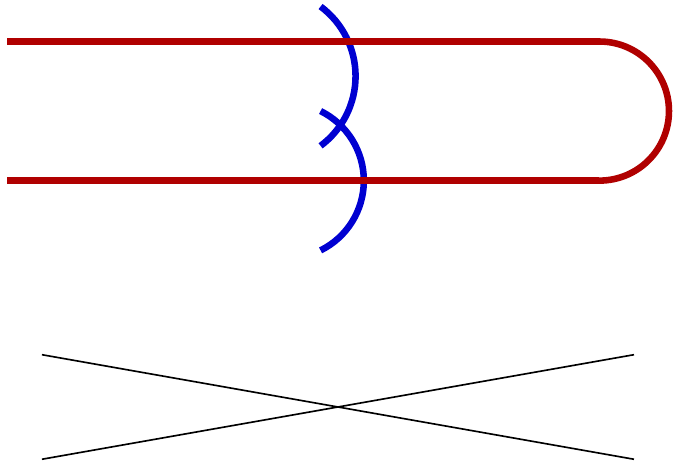
    \end{minipage}
  }
  \caption[Left:  
  charged hypermultiplet. Right: monodromy prevents the charge.]
  {Left: A localized hypermultiplet
    charged under a $U(1)$ from the difference of two sections. Right: A global monodromy preventing  the
     $U(1)$ gauge charge, resulting in a localized uncharged
    hypermultiplet.}
  \label{fig:U1monodromy}
\end{figure}
It is instructive to compare this hypermultiplet with a geometrically
similar hypermultiplet that arises \cite{MW} when there are two sections of the
fibration (instead of a single $2$-section).
On the right-hand side of \autoref{fig:U1monodromy}, we
have drawn the $2$-section and how it intersects the $I_2$ fiber over
the node in the base. Locally, this is indistinguishable from two
distinct sections generating a rank-one Mordell-Weil group (left
figure). The difference is only visible globally: either the two local
sections stay separate globally or they meet and are exchanged at a
ramification point. 
In either case there is a massless hypermultiplet,
either charged under the $U(1)$ determined by the difference of the
sections, or uncharged 
in the case of a $2$-section.

Note that the $U(1)$ charge of a hypermultiplet changes sign if we
exchange the role of the zero-section and the generating section of
the Mordell-Weil group.
Therefore, we can no longer assign $U(1)$
charges to the localized hypermultiplets 
if the global
monodromy breaks the gauge group.\footnote{Similar issues are discussed in 
\cite{box-graphs}.}

%%% Local Variables:
%%% TeX-master: "Main"
%%% eval: (TeX-PDF-mode 1)
%%% End:

\section[Triality of {\boldmath $E_6$} and Monodromies]%
[Triality of E6 and Monodromies]%
{Triality of {\boldmath $E_6$} and Monodromies}
\label{sec:E6}

\subsection{Unbroken Gauge Group}
\label{sec:no_monodromy}

In this section we will take a closer look at fibrations with a
Kodaira fiber of type $IV^*$, which translates into an $E_6$ gauge group
possibly broken by monodromies. In fact, we will find new monodromy
effects in \autoref{sec:3x_monodromy}. But for the sake of a coherent
presentation we will first review the two known classes, which are
known as split and non-split case \cite{Bershadsky:1996nh}.

To start, let us look at the unique split $E_6$ toric elliptic
fibration over $\IP^2$, that is, the model whose gauge group is
unbroken $E_6$ and nothing else. The ambient space is the toric
variety with polytope $\nabla_1$ whose points are
\begin{equation}
  \begin{array}{|c|c|ccccccc|ccc|ccccc|}
    \hline
    u & v & w_0 & w_1 & w_2 & w_3 & w_4 & w_5 & w_6 &
    f_{0}&f_{1}&f_{2} &&&&&
    \\ \hline
    6 & 0 & 3 & 3 & 2 & 1 & 3 & 1 & 0 & 3 & -1 & 0 &
    2 & 2 & 1 & 1 & 1 \\
    4 & 0 & 2 & 2 & 1 & 1 & 2 & 0 & 0 & 2 & 0 & -1 &
    1 & 1 & 1 & 1 & 0 \\
    -1 & 0 & 3 & 2 & 2 & 2 & 1 & 1 & 1 & 0 & 0 & 0 &
    0 & 1 & 0 & 1 & 0 \\
    -1 & 1 & 0 & 0 & 0 & 0 & 0 & 0 & 0 & 0 & 0 & 0 &
    0 & 0 & 0 & 0 & 0 \\
    \hline
    \multicolumn{9}{|c|}{\text{three tops}} & 
    \multicolumn{3}{c|}{\text{fiber}} & 
    \multicolumn{5}{c|}{\text{facet interior}}
    \\
    \hline
  \end{array}
\end{equation}
and which is fibered over $\IP^2$ via the projection to the last two
coordinates. In terms of homogeneous coordinates, this is the map
\begin{figure}[htbp]
  \centering
  \raisebox{-0.5\height}{
    \includegraphics{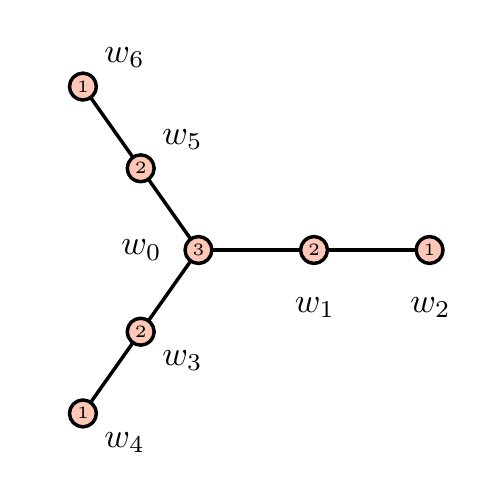}}%
  \raisebox{-0.5\height}{
    \includegraphics[width=0.7\textwidth]{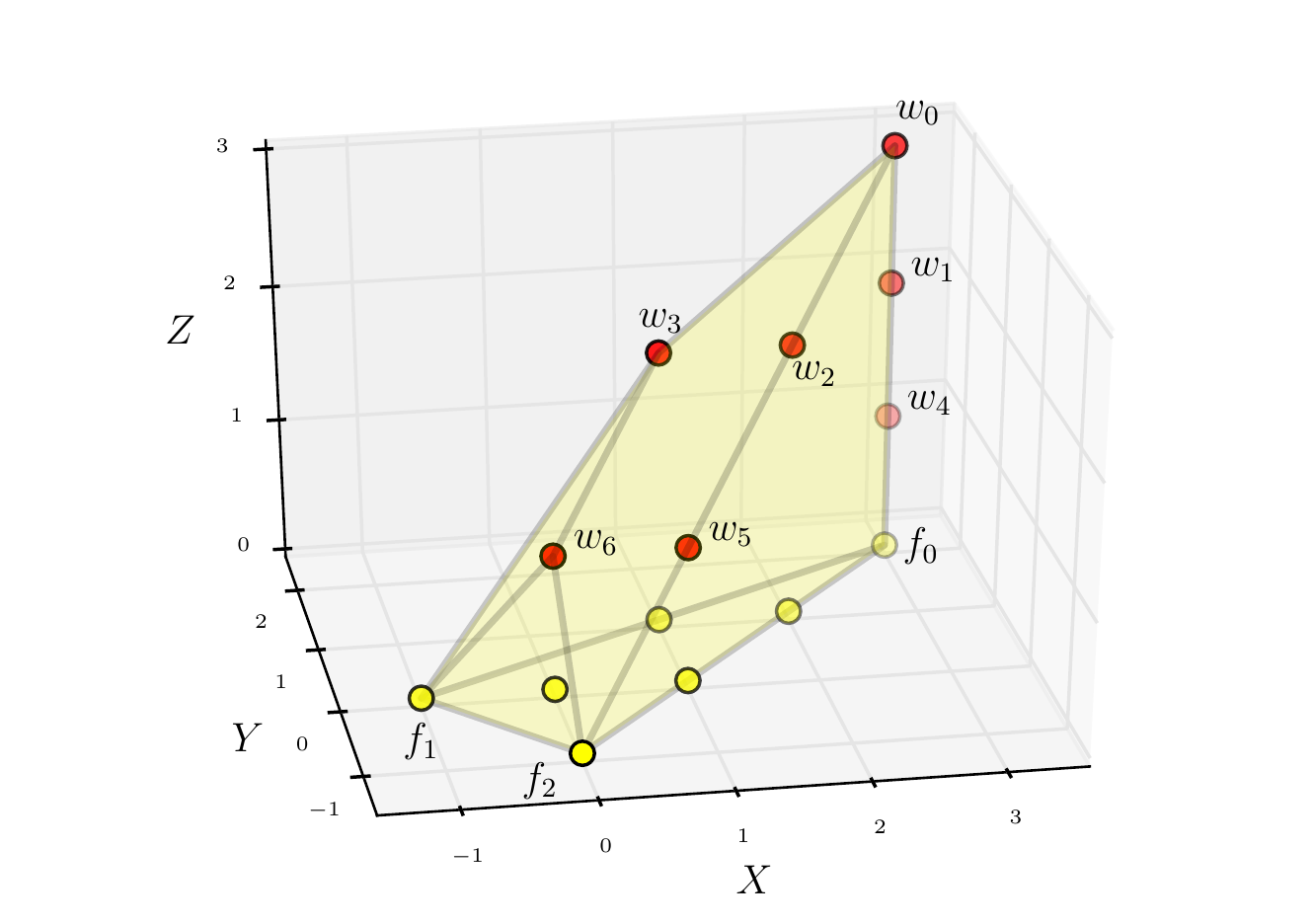}}
  \caption{The split $E_6$ top and associated Dynkin diagram.}
  \label{fig:E6split}
\end{figure}
\begin{equation}
  \pi: X_{\nabla_1} \to \IP^2
  ,\quad
  [u:v:w_0:\cdots:w_6:f_0:f_1:f_2] \mapsto
  \big[u:v:w_0^3 w_1^2 w_2^2 w_3^2 w_4 w_5 w_6 \big]
\end{equation}
In the case at hand, the $w$-top and Dynkin diagram are shown in
\autoref{fig:E6split}, which confirms that the gauge group is $E_6$
with no monodromy. Since the fiber polygon is that of $\IP^2[1,2,3]$,
we are in the favorable case where a Calabi-Yau hypersurface coincides with its
Weierstrass model. There is a toric section $V(f_0)$ meeting the fiber
component $V(w_4)\cap F$ in a point~\cite{Grassi:2012qw}, a toric
2-section $V(f_2)$ meeting $V(w_5)\cap F$ and $V(w_6)\cap F$ in one
point each, and a toric 3-section $V(f_3)$ meeting $V(w_6)\cap F$ in a
single point and meeting $V(w_3)\cap F$ in a point with multiplicity
two (for a total of 3 points). It is straightforward to check that the
Tate algorithm~\cite{MR0393039, Bershadsky:1996nh, Katz:2011qp} and
monodromy equation~\cite{Grassi:2011hq} agrees with the above
identification of the gauge group.

The elliptic fiber degenerates further over a number of points in the
base. The most basic degeneration is where the vanishing degree $(\deg
f,\ \deg g,\ \deg \Delta)$ of the Weierstrass coefficients and
discriminant is enhanced from $(3,4,8)$ to $(3,5,9)$. In general the
degeneration encodes charged matter content of the 6-d theory, and
this particular one yields a $\Rep{27}$ of $E_6$, and requires an
intersection of the $IV^*$ discriminant component and the $I_1$
discriminant component with multiplicity $4$. Hence, by degree
counting, one expects that the toric $IV^*$ discriminant component
intersect the $I_1$ discriminant component in $7$ points of
multiplicity $4$. A numerical analysis of a general toric hypersurface
equation confirms this. To summarize, the generic Calabi-Yau
hypersurface $Y_1 \subset X_{\nabla_1}$ has Hodge numbers
$h^{1,1}(Y_1)=8$, $h^{2,1}(Y_1)=161$. The 6-d F-theory
compactification is a $E_6$ gauge theory with $7\times \Rep{27}$. This
satisfies the gauge and gravitational anomaly cancellation conditions
\begin{equation}
  \begin{gathered}
    H_u = h^{2,1} + 1 = 162 ,\quad H_c = 7 \cdot 27 ,\quad V = 78
    \\
    H_u+H_c-V = 273 
    ,\quad 
    18 b = 7\cdot 6 - 24 
    ,\quad
    3 b^2 = 7\cdot 3 - 18
  \end{gathered}
\end{equation}
for $b=1$.

\subsection{Alternative Gauge Groups}
\label{sec:D4andA23}

Strictly speaking, the smooth Calabi-Yau threefold constructed in
\autoref{sec:no_monodromy} does not yield any gauge
interactions. Instead, one must contract some irreducible fiber
components. The ensuing singularity is what is responsible for the 5-d
gauge group, from which we infer the 6-d gauge group.  (The compactification
to 5-d gives the Coulomb branch of the 6-d theory, which may exhibit a
variety of unbroken subgroups, all containing the Cartan subgroup.)
The conventional way in which the 6-d group was determined was to generate
a 5-d gauge group by contracting all
irreducible fiber components except the one meeting the section,
which produced a maximal group.  The inference that this must coincide
with the 6-d group is clear.

Note that when there is more than one section a choice must be made,
but the groups obtained are all isomorphic to each other.  However,
when there is no section, the situation is different and it is possible
to produce different maximal 5-d gauge groups which are not contained in 
one another (somewhat analogous to the situation described in 
\cite{Witten:1997kz}).  We will see explicit examples of this shortly.

However, let us first just consider the $E_6$ theory from
\autoref{sec:no_monodromy}. We can, for example, shrink all components
except the three irreducible fiber components at the three ends of the
$E_6$ extended Dynkin diagram, even though the unique section passes
through only one of them. This amounts to the Levi type\footnote{That
  is, a branching rule corresponding to the removal of nodes from the
  (non-extended) Dynkin diagram.} branching rule
\begin{equation}
  \label{eq:E6branchD4}
  \begin{split}
    E_6 \supset&\; D_4
    , \\
    \Ad(E_6) = &\;
    \Ad(D_4) \oplus
    2 \big(\Rep{8}_v \oplus \Rep{8}_s \oplus \Rep{8}_c \big) 
    \oplus 2\times \Rep{1}
    , \\
    \Rep{27} = &\;
    \Rep{8}_v \oplus \Rep{8}_s \oplus \Rep{8}_c
    \oplus 3\times \Rep{1}
    .
  \end{split}
\end{equation}
The anomaly virtual representation restricts to
\begin{equation}
  H-V = 
  7\times \Rep{27} \ominus \Ad(E_6) =
  5 \big(\Rep{8}_v \oplus \Rep{8}_s \oplus \Rep{8}_c \big)  
  \oplus 19\times \Rep{1}
  \ominus \Ad(D_4)
  ,
\end{equation}
and we obtain a $D_4=SO(8)$ gauge theory with $5$ copies of vector,
spinor, and conjugate spinor as well as $H_u=h^{2,1}+1+19$ uncharged
hypers. Of course we recognize this as the Higgs mechanism: some of
the vector multiplets got massive, eating a hypermultiplet partner in
the process. In particular, the anomaly cancellation condition is
preserved.

By contrast, let us now shrink all irreducible fiber components except
the central one\footnote{Note that a section cannot pass though the
  central node: Its multiplicity is three, so only $(3n)$-sections can
  pass through it.} in the $E_6$ extended Dynkin diagram, see
\autoref{fig:E6split}. This is allowed by the geometry, that is, there
exists a particular triangulation of the face fan of the polyhedron
such that $V(w_0)$ is the only fibral divisor that does not vanish at
a particular face of the K\"ahler cone. Contracting two simply laced
nodes creates an $A_2$ singularity, so the resulting gauge group is
$A_2^3=SU(3)^3$. The corresponding branching rule is
\begin{equation}
  \label{eq:E6branchA23}
  \begin{split}
    E_6 \supset&\; SU(3)^3
    , \\
    \Ad(E_6) = &\;
    (\Rep{8},\Rep{1},\Rep{1}) \oplus 
    (\Rep{1},\Rep{8},\Rep{1}) \oplus 
    (\Rep{1},\Rep{1},\Rep{8}) \oplus
    (\Rep{3},\Rep{3},\Rep{3}) \oplus 
    (\barRep{3}, \barRep{3}, \barRep{3})
    , \\
    \Rep{27} = &\;
    (\Rep{3},\barRep{3},\Rep{1}) \oplus
    (\Rep{1},\Rep{3},\barRep{3}) \oplus
    (\barRep{3},\Rep{1},\Rep{3})
    .
  \end{split}
\end{equation}
This is an extended type branching rule, corresponding to the removal
of a node from the \emph{extended} Dynkin diagram. Note that the $E_6$
and $A_2^3$ theories stand on the same footing as both correspond to
the removal of a single node from the extended Dynkin diagram. We are
just picking a different compact subgroup of the affine $E_6$. In
particular, neither can be obtained by the Higgs mechanism from the
other. Note that the 6-d theory obtained this way is not a standard
gauge theory as it contains exotic vector multiplets in the
$(\Rep{3},\Rep{3},\Rep{3})$ and $(\barRep{3}, \barRep{3}, \barRep{3})$
representation in addition to the gauge multiplets.

\subsection{The Non-Split Case}
\label{sec:2x_monodromy}

\begin{figure}[htbp]
  \centering
  \raisebox{-0.5\height}{
    \includegraphics{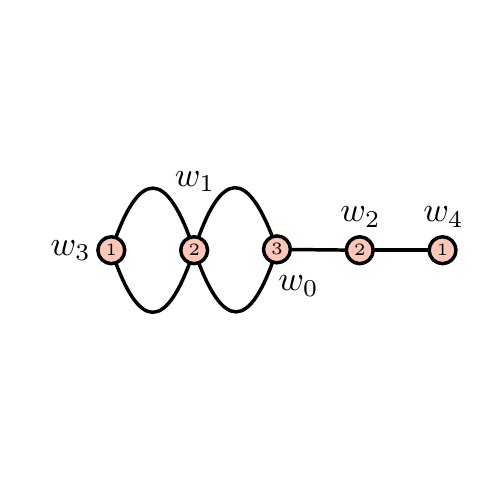}}%
  \raisebox{-0.5\height}{
    \includegraphics[width=0.7\textwidth]{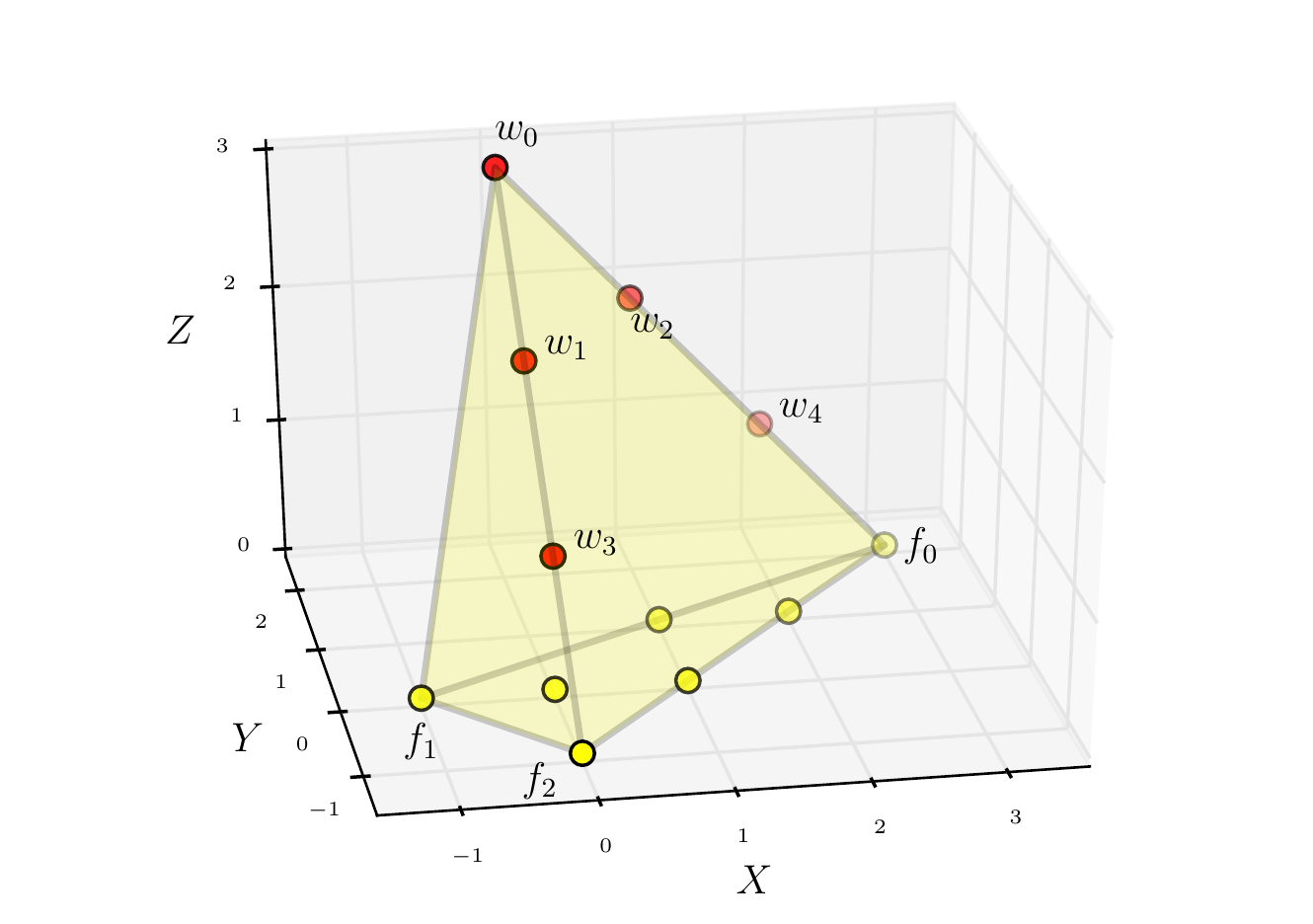}}
  \caption[$F_4$ top with $\Z_2$ monodromy and Dynkin diagram.]
  {The $F_4$ top with $\Z_2$ monodromy and associated
    contracted Dynkin diagram.}
  \label{fig:E6_2x_monodromy}
\end{figure}

We now proceed to the so-called \emph{non-split} $IV^*$, that is, a
$IV^*$ Kodaira fiber with a $\Z_2$ monodromy exchanging two of the
three legs~\cite{DegeratuWendland}. 
This is interpreted as $E_6$ broken to $F_4$ by the
monodromy. The toric ambient space corresponds to the polytope
$\nabla_2$ with points
\begin{equation}
  \begin{array}{|c|c|ccccc|ccc|ccccc|}
    \hline
    u & v & w_0 & w_1 & w_2 & w_3 & w_4 &
    f_{0}&f_{1}&f_{2} &&&&&
    \\ \hline
    0 & 7 & 0 & 0 & 1 & 0 & 2 & 3 & -1 & 0
    & 0 & 1 & 1 & 1 & 2 \\
    0 & 4 & 2 & 1 & 2 & 0 & 2 & 2 & 0 & -1
    & 1 & 0 & 1 & 1 & 1 \\
    -1 & 0 & 3 & 2 & 2 & 1 & 1 & 0 & 0 & 0
    & 1 & 0 & 0 & 1 & 0 \\
    -1 & 1 & 0 & 0 & 0 & 0 & 0 & 0 & 0 & 0
    & 0 & 0 & 0 & 0 & 0 \\
    \hline
    \multicolumn{7}{|c|}{\text{three tops}} & 
    \multicolumn{3}{c|}{\text{fiber}} & 
    \multicolumn{5}{c|}{\text{facet interior}}
    \\
    \hline
  \end{array}
\end{equation}
The coordinates are again chosen such that the only non-trivial
discriminant component comes from the $w$-top, which is depicted in
\autoref{fig:E6_2x_monodromy}.
The toric section $V(f_0)$ intersects the irreducible fiber component
$V(w_4)\cap F$ in a single point, and the toric $3$-section $V(f_2)$
intersects the irreducible fiber component $V(w_0)\cap F$ in a point
with multiplicity $3$. The toric $2$-section $V(f_2)$ intersects the
fiber component $V(w_3)\cap F$ in two points. Since the latter has
multiplicity one, the fiber component $V(w_3)\cap F$ must consist of
two irreducible components. This explains why the Dynkin diagram is
contracted as on the left hand side of \autoref{fig:E6_2x_monodromy}.

The standard choice in contracting the fiber components is to contract
every one except $V(w_4)$. This leads to a $F_4$ gauge theory, namely
$E_6$ broken by the $\Z_2$ monodromy. As for the matter content, one
again expects the most simple degeneration of the Weierstrass model
where the vanishing degree $(\deg f,\ \deg g,\ \deg \Delta)$ jumps
from $(3,4,8)$ to $(3,5,9)$. If one allows non-trivial monodromy, this
requires a double intersection of the $IV^*$ and $I_1$ divisor
components. Hence, by degree counting, there should be $14$ such
points and a numerical analysis of the Weierstrass hypersurface
equation indeed confirms this. Unlike in the case without monodromy,
there is no curve stuck over these codimension-two points in the
base. Instead, the matter comes from one-parameter families of curves
over the $IV^*$ discriminant locus and is determined by the Euler
characteristic of their moduli space. In particular, the moduli space
here is the base $\IP^1$ branched at $14$ points, that is, a Riemann
surface of genus $6$. To summarize, the generic Calabi-Yau
hypersurface $Y_2\subset X_{\nabla_2}$ has Hodge numbers
$h^{1,1}(Y_2)=6$, $h^{2,1}(Y_2)=180$. The 6-d F-theory
compactification is a $F_4$ gauge theory with $6\times \Rep{26}$
matter hypermultiplets.\footnote{This matches the restricting of the
  anomaly virtual representation $H-V$ under the $E_6 \supset F_4$
  branching rule. Since the branching rule is of symmetric type, this
  breaking is not the traditional Higgs mechanism.} This satisfies the
gauge and gravitational anomaly cancellation conditions\footnote{Note
  that the \emph{charged dimension}~\cite{Grassi:2000we,
    Grassi:2011hq} of $\Rep{26}$ is $24$.  We review its definition in
  \autoref{sec:sage}. In order to count a charged hyper as a complete
  $\Rep{26}$, we have to also remove two unchanged hypers.}
\begin{equation}
  \label{eq:F4anomaly}
  \begin{gathered}
    H_u = h^{2,1} + 1 - 6\cdot 2 = 169 
    ,\quad 
    H_c = 6 \cdot 26 ,\quad V = 52
    \\
    H_u+H_c-V = 273 
    ,\quad 
    18 b = 6\cdot 6 - 18 
    ,\quad
    3 b^2 = 6\cdot 3 - 15
  \end{gathered}
\end{equation}
for $b=1$.

\subsection{Starting at the Other End}
\label{sec:other_end}

\begin{figure}[tbp]
  \centering
  \raisebox{-0.5\height}{
    \includegraphics[width=0.35\linewidth]{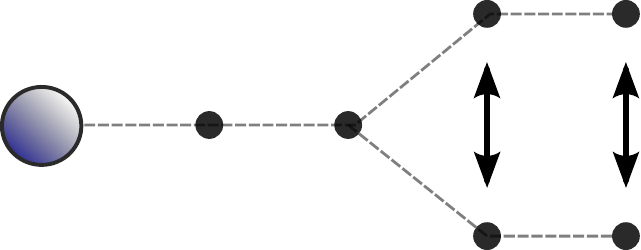}
  }
  \hspace{2cm}
  \raisebox{-0.5\height}{
    \includegraphics[width=0.35\linewidth]{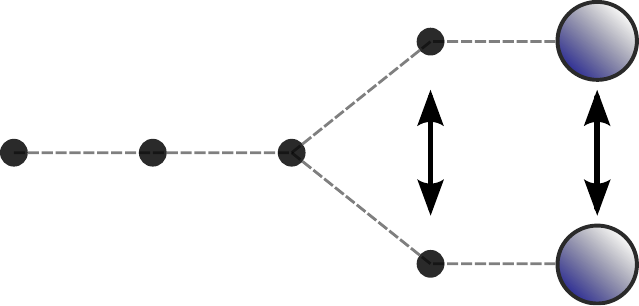}
  }
  \caption[Different contractions of the $IV^*$ with monodromy.]
  {Different contractions of the $IV^*$ Kodaira fiber with
    $\Z_2$ monodromy (vertical arrows). Left: contract all components
    not intersecting the given section $\Rightarrow F_4$. Right:
    contract everything except the rightmost monodromy orbit
    $\Rightarrow B_4$}
  \label{fig:IVstar_Z2_contractions}
\end{figure}

Instead of picking the fiber component intersecting the given section,
we can also pick the two fiber components, exchanged by the monodromy,
at the other end of the contracted Dynkin diagram (see
\autoref{fig:IVstar_Z2_contractions}). Shrinking all the complementary
fiber components, we arrive at a $B_4$ gauge theory. We can use the
extended-type $F_4\supset B_4$ branching rule
\begin{equation}
  \label{eq:F4branchB4}
  \begin{split}
    F_4 \supset&\; B_4
    , \\
    \Ad(F_4) = &\;
    \Ad(B_4) \oplus \Rep{16}
    , \\
    \Rep{26} = &\;
    \Rep{16} \oplus \Rep{9} \oplus \Rep{1}
  \end{split}
\end{equation}
to convert one into the other. To actually break the gauge group to
$B_4$, the $\Rep{16}$ vector multiplet pairs up with a corresponding
hypermultiplet and becomes massive. Hence we obtain a $B_4$ gauge
theory with $5\times\Rep{16}$ and $6\times \Rep{9}$ charged
hypermultiplets. The charged dimension of $\Rep{9}$ is only $8$, so we
have to subtract $6$ uncharged hypermultiplets to avoid
overcounting. The resulting matter content cancels the gauge and
gravity anomalies
\begin{equation}
  \label{eq:B4_gauge}
  \begin{gathered}
    H_u 
    = h^{2,1} + 1 - 6\times 1 
    = 175 
    , \quad
    H_c = 5 \cdot 16 + 6 \cdot 9 
    ,\quad 
    V = 36
    \\
    H_u + H_c - V = 273
    ,
    \\
    18 b = 5 \cdot 4 + 6\cdot 2 - 14 
    ,\quad
    0 = 5 \cdot (-2) + 6\cdot 2 - 2
    ,\quad
    3 b^2 = 5 \cdot 3 + 6\cdot 0 - 12
  \end{gathered}
\end{equation}
for $b=1$.

More systematically, we can derive the hyper and vector multiplet
count from the geometry of the elliptic fibration. Since identifying
the correct branching rule will be somewhat tricky in an example that
we encounter later on, let us walk through the more pedestrian
argument. In the case with monodromy, note that there are no isolated
codimension-two curves stuck over a point in the base and the proper
way to count the non-isolated curves is by the genus of their moduli
space. The trick is to identify all (possibly reducible) curves, count
the number of massless fields supported on them, and then reassemble
the component fields into gauge multiplets. For purposes of the
exposition, let us stick to an $IV^*$ Kodaira fiber whose irreducible
fiber components correspond to the simple $\Etilde_6$ roots. The roots
are generated by the simple roots, and we henceforth identify
\begin{sidewaystable}[p]
  \centering
  \hspace{-1cm}
  \begin{tabular}{c@{$\quad ($}r@{,\;}r@{,\;}r@{,\;}r@{$)\quad$}ccc}
    % shrink and symmetric
    \multicolumn{1}{c}{$\Gamma$-orbit} & 
    \multicolumn{4}{c}{$B_4$-weight} & 
    $g'$ & H & V
    \\ \hline
    \smallESixAffineLabels0122321 & $-2$&$1$&$0$&$0$ & $0$ & $0$ & $1$ \\
    \smallESixAffineLabels0112321 & $-1$&$-1$&$1$&$0$  & $0$ & $0$ & $1$ \\
    \smallESixAffineLabels0112221 & $-1$&$0$&$-1$&$1$  & $0$ & $0$ & $1$ \\
    \smallESixAffineLabels0111211 & $-1$&$0$&$1$&$-1$  & $0$ & $0$ & $1$ \\
    \smallESixAffineLabels0011210 & $-1$&$0$&$1$&$0$  & $0$ & $0$ & $1$ \\
    \smallESixAffineLabels0111111 & $-1$&$1$&$-1$&$0$  & $0$ & $0$ & $1$ \\
    \smallESixAffineLabels0011110 & $-1$&$1$&$-1$&$1$  & $0$ & $0$ & $1$ \\
    \smallESixAffineLabels0010100 & $-1$&$1$&$1$&$-1$  & $0$ & $0$ & $1$ \\
    \smallESixAffineLabels0010000 & $-1$&$2$&$-1$&$0$  & $0$ & $0$ & $1$ \\
    \smallESixAffineLabels0101111 & $0$&$-1$&$0$&$0$  & $0$ & $0$ & $1$ \\
    \smallESixAffineLabels0001110 & $0$&$-1$&$0$&$1$  & $0$ & $0$ & $1$ \\
    \smallESixAffineLabels0000100 & $0$&$-1$&$2$&$-1$  & $0$ & $0$ & $1$ \\
    \smallESixAffineLabels1122221 & $0$&$1$&$-2$&$1$  & $0$ & $0$ & $1$ \\
    \smallESixAffineLabels1121211 & $0$&$1$&$0$&$-1$  & $0$ & $0$ & $1$ \\
    \smallESixAffineLabels1021210 & $0$&$1$&$0$&$0$  & $0$ & $0$ & $1$ \\
    \smallESixAffineLabels1112321 & $1$&$-2$&$1$&$0$  & $0$ & $0$ & $1$ \\
    \smallESixAffineLabels1112221 & $1$&$-1$&$-1$&$1$  & $0$ & $0$ & $1$ \\
    \smallESixAffineLabels1111211 & $1$&$-1$&$1$&$-1$  & $0$ & $0$ & $1$ \\
    \smallESixAffineLabels1011210 & $1$&$-1$&$1$&$0$  & $0$ & $0$ & $1$ \\
    \smallESixAffineLabels1111111 & $1$&$0$&$-1$&$0$  & $0$ & $0$ & $1$ \\
    \smallESixAffineLabels1011110 & $1$&$0$&$-1$&$1$  & $0$ & $0$ & $1$ \\
    \smallESixAffineLabels1010100 & $1$&$0$&$1$&$-1$  & $0$ & $0$ & $1$ \\
    \smallESixAffineLabels1010000 & $1$&$1$&$-1$&$0$  & $0$ & $0$ & $1$ \\
    \smallESixAffineLabels1000000 & $2$&$-1$&$0$&$0$  & $0$ & $0$ & $1$ \\
  \end{tabular}
  \hspace{5mm}
  \begin{tabular}{c@{$\quad ($}r@{,\;}r@{,\;}r@{,\;}r@{$)\quad$}ccc}
    \multicolumn{1}{c}{$\Gamma$-orbit} & 
    \multicolumn{4}{c}{$B_4$-weight} & 
    $g'$ & H & V
    \\ \hline
    % shrink and non-symmetric
    \smallESixAffineLabels0111221 ~ \smallESixAffineLabels0112211
    & $-1$&$0$&$0$&$0$  & $6$ & $6$ & $1$ \\
    \smallESixAffineLabels0010110 ~ \smallESixAffineLabels0011100
    & $-1$&$1$&$0$&$0$  & $6$ & $6$ & $1$ \\
    \smallESixAffineLabels0000110 ~ \smallESixAffineLabels0001100
    & $0$&$-1$&$1$&$0$  & $6$ & $6$ & $1$ \\
    \smallESixAffineLabels0000010 ~ \smallESixAffineLabels0001000
    & $0$&$0$&$-1$&$1$  & $6$ & $6$ & $1$ \\
    \smallESixAffineLabels1121321 ~ \smallESixAffineLabels1122311
    & $0$&$0$&$1$&$-1$  & $6$ & $6$ & $1$ \\
    \smallESixAffineLabels1121221 ~ \smallESixAffineLabels1122211
    & $0$&$1$&$-1$&$0$  & $6$ & $6$ & $1$ \\
    \smallESixAffineLabels1111221 ~ \smallESixAffineLabels1112211
    & $1$&$-1$&$0$&$0$  & $6$ & $6$ & $1$ \\
    \smallESixAffineLabels1010110 ~ \smallESixAffineLabels1011100
    & $1$&$0$&$0$&$0$  & $6$ & $6$ & $1$ \\
    % finite size non-symmetric
    \smallESixAffineLabels0011221 ~ \smallESixAffineLabels0112210
    & $-1$&$0$&$0$&$\tfrac{1}{2}$  & $6$ & $5$ & $0$ \\
    \smallESixAffineLabels0011211 ~ \smallESixAffineLabels0111210
    & $-1$&$0$&$1$&$-\tfrac{1}{2}$  & $6$ & $5$ & $0$ \\
    \smallESixAffineLabels0011111 ~ \smallESixAffineLabels0111110
    & $-1$&$1$&$-1$&$\tfrac{1}{2}$  & $6$ & $5$ & $0$ \\
    \smallESixAffineLabels0010111 ~ \smallESixAffineLabels0111100
    & $-1$&$1$&$0$&$-\tfrac{1}{2}$  & $6$ & $5$ & $0$ \\
    \smallESixAffineLabels0001111 ~ \smallESixAffineLabels0101110
    & $0$&$-1$&$0$&$\tfrac{1}{2}$  & $6$ & $5$ & $0$ \\
    \smallESixAffineLabels0000111 ~ \smallESixAffineLabels0101100
    & $0$&$-1$&$1$&$-\tfrac{1}{2}$  & $6$ & $5$ & $0$ \\
    \smallESixAffineLabels0000011 ~ \smallESixAffineLabels0101000
    & $0$&$0$&$-1$&$\tfrac{1}{2}$  & $6$ & $5$ & $0$ \\
    \smallESixAffineLabels0000001 ~ \smallESixAffineLabels0100000
    & $0$&$0$&$0$&$-\tfrac{1}{2}$  & $6$ & $5$ & $0$ \\
    \smallESixAffineLabels1022321 ~ \smallESixAffineLabels1122320
    & $0$&$0$&$0$&$\tfrac{1}{2}$  & $6$ & $5$ & $0$ \\
    \smallESixAffineLabels1021321 ~ \smallESixAffineLabels1122310
    & $0$&$0$&$1$&$-\tfrac{1}{2}$  & $6$ & $5$ & $0$ \\
    \smallESixAffineLabels1021221 ~ \smallESixAffineLabels1122210
    & $0$&$1$&$-1$&$\tfrac{1}{2}$  & $6$ & $5$ & $0$ \\
    \smallESixAffineLabels1021211 ~ \smallESixAffineLabels1121210
    & $0$&$1$&$0$&$-\tfrac{1}{2}$  & $6$ & $5$ & $0$ \\
    \smallESixAffineLabels1011221 ~ \smallESixAffineLabels1112210
    & $1$&$-1$&$0$&$\tfrac{1}{2}$  & $6$ & $5$ & $0$ \\
    \smallESixAffineLabels1011211 ~ \smallESixAffineLabels1111210
    & $1$&$-1$&$1$&$-\tfrac{1}{2}$  & $6$ & $5$ & $0$ \\
    \smallESixAffineLabels1011111 ~ \smallESixAffineLabels1111110
    & $1$&$0$&$-1$&$\tfrac{1}{2}$  & $6$ & $5$ & $0$ \\
    \smallESixAffineLabels1010111 ~ \smallESixAffineLabels1111100
    & $1$&$0$&$0$&$-\tfrac{1}{2}$  & $6$ & $5$ & $0$ \\
  \end{tabular}
  \hspace{-2cm}
  \caption[$B_4$-weights of curves in the $IV^*$ Kodaira fiber with
  $\Z_2$ monodromy.]{The $\Z_2$-orbits of the $72$ curves, $B_4$
    weight, genus of the moduli space, and the resulting number of
    hyper and vector multiplets. The weight under $B_4$ is the
    intersection product with $C_4$, $C_1$, $C_0$, and
    $\tfrac{1}{2}(C_2+C_3)$.}
  \label{tab:E6B4matter}
\end{sidewaystable}
\begin{itemize}
\item The $72$ roots $\pm \alpha_i$ of $E_6$,
\item The affine roots 
  \begin{math}
    \bigcup \big\{ (\alpha; 0) ,\ (\theta-\alpha; 1) \big\}
  \end{math}
  where $\theta$ is the highest root of $E_6$,
\item Curves
  \begin{equation}
    C = \sum a_i C_i
    =
    \left(
      \ESixAffineLabels{a_4}{a_5}{a_1}{a_2}{a_0}{a_3}{a_6}
    \right)
  \end{equation}
    of self-intersection $C\cdot C = -2$ in $F$.
\end{itemize}
Clearly, the positive roots $\alpha_i$ correspond to the affine roots
$(\alpha_i; 0)$ which correspond to curves not wrapping the extended
node in the Dynkin diagram. Flipping the sign of $\alpha_i$
corresponds to $\hat{\alpha}_i\mapsto (\theta,1)-\hat{\alpha}_i$ which
corresponds to $C\mapsto F-C$.

We are interested in the case where the gauge group is broken both by
leaving some irreducible fiber components at finite size and by a
monodromy group $\Gamma$ acting by permutation on the affine roots. A
fiber component $C$ can be moved along the discriminant, and we take
the moduli space to be a curve of genus $g'(\mathcal{M}_C)$. We say
that a curve shrinks if either $C$ or $F-C$ has zero volume, that is,
does not wrap any irreducible fiber component chosen to have finite
size.
\begin{itemize}
\item Each monodromy orbit $\Gamma C$ of a shrinking curve yields
  $g'(\mathcal{M}_C)$ hypermultiplets and $1$ vector multiplet.
\item Each monodromy orbit $\Gamma C$ of a finite-size curve yields
  $g'(\mathcal{M}_C)-1$ hypermultiplets and no vector multiplets.
\end{itemize}
In the $B_4$ example, we distinguish two curve types: If the curve is
fixed by the monodromy, then its moduli space equals the corresponding
discriminant component (which is a $\CP^1$). If the fiber component is
exchanged by monodromy, then it is a branched double cover of $\CP^1$.
The number of branch points can be read off from the Weierstrass
model, and we find $14$. Hence, the moduli space of curves that are
not fixed under the monodromy action is a Riemann surface of genus
$g'=6$. The roots are tabulated in \autoref{tab:E6B4matter}. We see
that, indeed, the spectrum consists of a vector multiplet in the
adjoint of $B_4$ as well as $5\times\Rep{16}$ and $6\times \Rep{9}$
hypermultiplets. Finally, we can also systematically count the number
of uncharged hypermultiplets. The weight zero subspace of the adjoint
of $E_8$ is $6$, which means that this many complex structure moduli
are actually frozen by requiring a $IV^*$ Kodaira fiber. This is
already accounted for in the gravitational anomaly where the complete
$V$ including its weight zero subspace is subtracted from the number
of complex structure moduli. To assemble the components in
\autoref{tab:E6B4matter} into complete $B_4$ representations, we need
$4$ uncharged fields for the adjoint and one uncharged field for the
$\Rep{9}$. The adjoint field is the gauge vector multiplet, and its
weight zero subspace is already accounted for. Therefore, the
remaining number of $B_4$-uncharged hypermultiplets equals
\begin{equation}
  H_u 
  = h^{2,1} + 1 - 6\cdot 1 
  = 175 
  \quad \Rightarrow \quad
  H_u + H_c - V = 273
  .
\end{equation}

\subsection{A Novel Monodromy Effect}
\label{sec:3x_monodromy}

The third model with a $IV^*$ Kodaira fiber under consideration will
have a different fiber ambient space, namely $\IP^2$ instead of
$\IP^2[1,2,3]$. The effect of this change is that the toric ambient
space no longer forces the fibration to have a section but only
$3$-sections. Explicitly, we consider the polytope $\nabla_3$ with
points
\begin{equation}
  \begin{array}{|c|c|ccc|ccc|c|}
    \hline
    u & v & w_0 & w_1 & w_2 &
    f_{0}&f_{1}&f_{2} &
    \\ \hline
    0 & -1 & 1 & 1 & 1 & 1 & 0 & -1 & 0 \\
    0 & 1 & 0 & 0 & 0 & 0 & 1 & -1 & 0 \\
    -1 & 0 & 3 & 2 & 1 & 0 & 0 & 0 & 1 \\
    -1 & 1 & 0 & 0 & 0 & 0 & 0 & 0 & 0 \\
    \hline
    \multicolumn{5}{|c|}{\text{three tops}} & 
    \multicolumn{3}{c|}{\text{fiber}} & 
    \multicolumn{1}{c|}{\text{facet interior}}
    \\
    \hline
  \end{array}
  \label{eq:E6_3x_polytope}
\end{equation}
where again the only non-trivial discriminant component comes from the
$w$-top. Each of $V(f_0)$, $V(f_1)$, and $V(f_2)$ is a toric
$3$-section. From the intersection numbers we conclude that,
restricted to a generic fiber $F$, $V(w_0)\cap F$ consists of a single
$\CP^1$ of multiplicity $3$, $V(w_1)\cap F$ consists of three disjoint
$\CP^1$ of multiplicity $2$, and $V(w_2)\cap F$ consists of three
disjoint $\CP^1$ of multiplicity $1$.
\begin{figure}[htbp]
  \centering
  \raisebox{-0.5\height}{
    \includegraphics{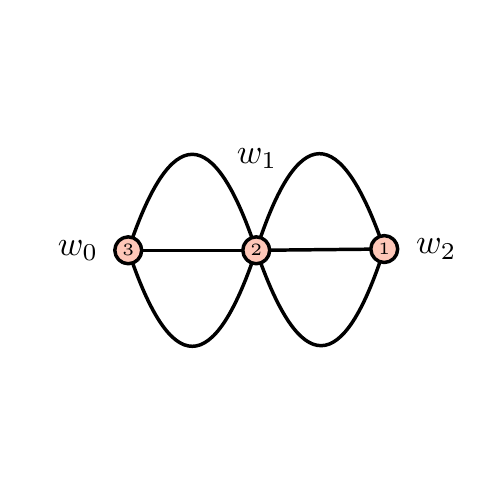}}%
  \raisebox{-0.5\height}{
    \includegraphics[width=0.7\textwidth]{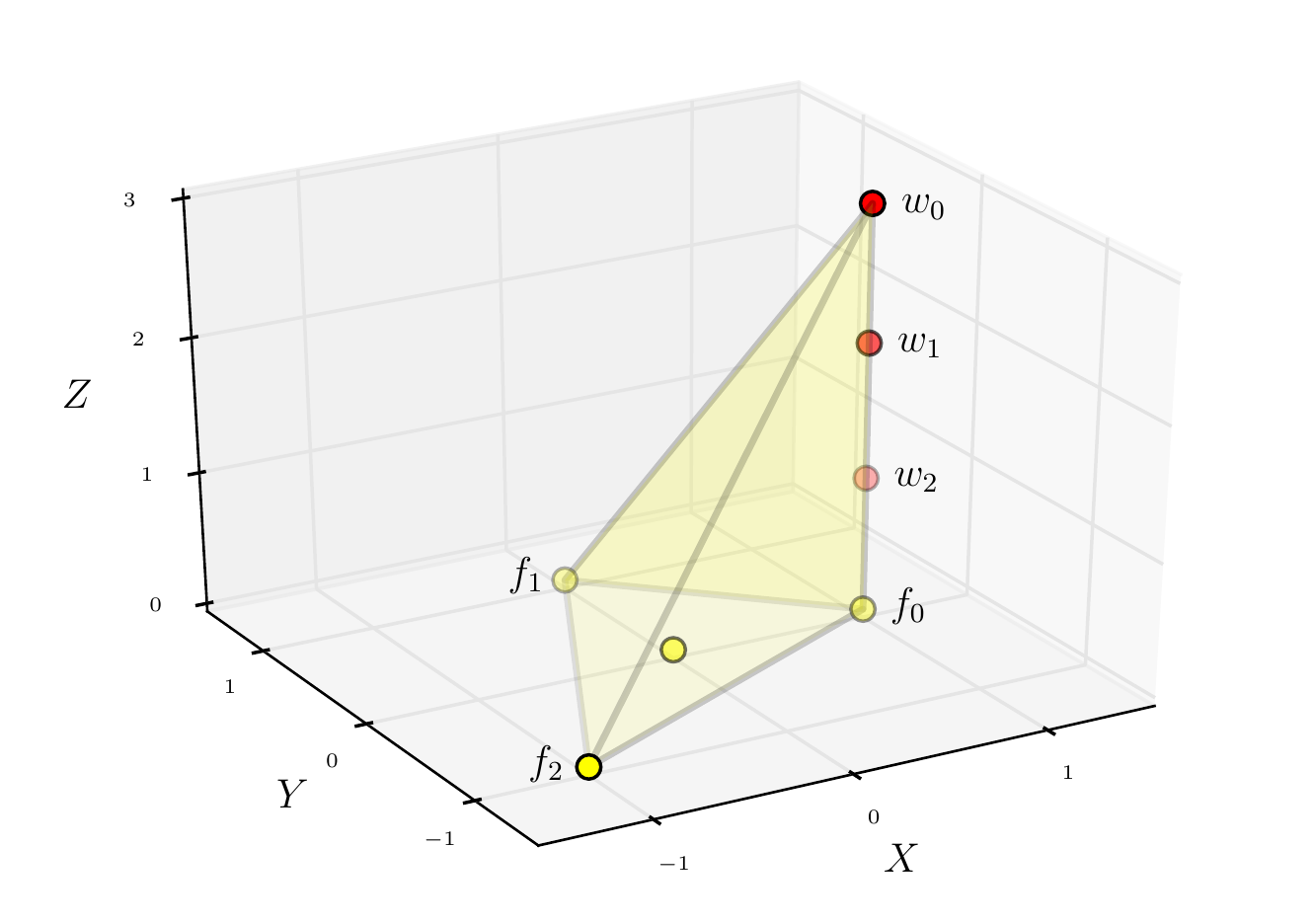}}
  \caption{The $G_2$ top with $\mathfrak{S}_3$-monodromy and associated Dynkin
    diagram.}
  \label{fig:E6_3x_monodromy}
\end{figure}
Therefore, the affine $E_6$ Dynkin diagram is folded as in
\autoref{fig:E6_3x_monodromy} into the ambient toric divisors. Each of
the three disconnected components of $V(w_2)\cap F$ intersects the
irreducible three-section $V(f_0)$ in a point. Hence, the monodromy
action freely permutes the three irreducible components in $V(w_2)\cap
F$. In particular, there cannot be a section (toric or not) in this
genus-one fibration as it would unambiguously mark one of the three
components.

As discussed in \autoref{sec:hyper_count}, the lack of a section
allows for nodes in the $I_1$ discriminant component. In fact, a
numerical analysis of a generic Calabi-Yau hypersurface finds that
there are $108$ nodes in the $I_1$ discriminant component, all of
which are away from the $IV^*$ discriminant component. This
contributes an extra $108$ uncharged hypermultiplets localized in
codimension-two over the base.
\begin{table}[htbp]
  \centering
  \begin{tabular}{c@{$\quad ($}r@{,\;}r@{$)\quad$}ccc}
    % shrink and symmetric
    \multicolumn{1}{c}{$\Gamma$-orbit} & 
    \multicolumn{2}{c}{$G_2$-wt.} & 
    $g'(C)$ & H & V
    \\ \hline
    \smallESixAffineLabels1111211 & $-3$&$3$  & $0$ & $0$ & $1$ \\
    \smallESixAffineLabels0000100 & $-3$&$6$  & $0$ & $0$ & $1$ \\
    \smallESixAffineLabels1111111 & $0$&$-3$  & $0$ & $0$ & $1$ \\
    \smallESixAffineLabels0011210 & $0$&$3$  & $0$ & $0$ & $1$ \\
    \smallESixAffineLabels1122221 & $3$&$-6$  & $0$ & $0$ & $1$ \\
    \smallESixAffineLabels0011110 & $3$&$-3$  & $0$ & $0$ & $1$ \\
    \smallESixAffineLabels1112321
    \smallESixAffineLabels1121321
    \smallESixAffineLabels1122311
    & $-2$&$3$  & $5$ & $5$ & $1$ \\
    \smallESixAffineLabels1111221
    \smallESixAffineLabels1112211
    \smallESixAffineLabels1121211
    & $-1$&$0$  & $5$ & $5$ & $1$ \\
    \smallESixAffineLabels0000110
    \smallESixAffineLabels0001100
    \smallESixAffineLabels0010100
    & $-1$&$3$  & $5$ & $5$ & $1$ \\
    \smallESixAffineLabels1112221
    \smallESixAffineLabels1121221
    \smallESixAffineLabels1122211
    & $1$&$-3$  & $5$ & $5$ & $1$ \\
    \smallESixAffineLabels0001110
    \smallESixAffineLabels0010110
    \smallESixAffineLabels0011100
    & $1$&$0$  & $5$ & $5$ & $1$ \\
    \smallESixAffineLabels0000010
    \smallESixAffineLabels0001000 
    \smallESixAffineLabels0010000
    & $2$&$-3$  & $5$ & $5$ & $1$ \\
    \smallESixAffineLabels0000111
    \smallESixAffineLabels0101100
    \smallESixAffineLabels1010100
    & $-2$&$3$  & $5$ & $4$ & $0$ \\
    \smallESixAffineLabels0111211
    \smallESixAffineLabels1011211
    \smallESixAffineLabels1111210
    & $-2$&$3$  & $5$ & $4$ & $0$ \\
    \smallESixAffineLabels0000001
    \smallESixAffineLabels0100000
    \smallESixAffineLabels1000000
    & $-1$&$0$  & $5$ & $4$ & $0$ \\
    \smallESixAffineLabels0101111
    \smallESixAffineLabels1010111
    \smallESixAffineLabels1111100
    & $-1$&$0$  & $5$ & $4$ & $0$ \\
    \smallESixAffineLabels0011211
    \smallESixAffineLabels0111210
    \smallESixAffineLabels1011210
    & $-1$&$3$  & $5$ & $4$ & $0$ \\
    \smallESixAffineLabels0112321
    \smallESixAffineLabels1021321
    \smallESixAffineLabels1122310
    & $-1$&$3$  & $5$ & $4$ & $0$ \\
    \smallESixAffineLabels0000011
    \smallESixAffineLabels0101000
    \smallESixAffineLabels1010000
    & $1$&$-3$  & $5$ & $4$ & $0$ \\
    \smallESixAffineLabels0111111
    \smallESixAffineLabels1011111
    \smallESixAffineLabels1111110
    & $1$&$-3$  & $5$ & $4$ & $0$ \\
    \smallESixAffineLabels0011221
    \smallESixAffineLabels0112210
    \smallESixAffineLabels1021210
    & $1$&$0$  & $5$ & $4$ & $0$ \\
    \smallESixAffineLabels0122321
    \smallESixAffineLabels1022321
    \smallESixAffineLabels1122320
    & $1$&$0$  & $5$ & $4$ & $0$ \\
    \smallESixAffineLabels0011111 
    \smallESixAffineLabels0111110 
    \smallESixAffineLabels1011110 
    & $2$&$-3$  & $5$ & $4$ & $0$ \\
    \smallESixAffineLabels0112221 
    \smallESixAffineLabels1021221
    \smallESixAffineLabels1122210
    & $2$&$-3$  & $5$ & $4$ & $0$ 
    \\[1ex]
    \begin{tabular}{c}
      \smallESixAffineLabels0001111
      \smallESixAffineLabels0010111
      \smallESixAffineLabels0101110 
      \\[-1ex]
      \smallESixAffineLabels0111100 
      \smallESixAffineLabels1010110 
      \smallESixAffineLabels1011100 
    \end{tabular}
    & $0$&$0$  & $16$ & $15$ & $0$ 
    \\[2ex]
    \begin{tabular}{c}
      \smallESixAffineLabels0111221 
      \smallESixAffineLabels0112211
      \smallESixAffineLabels1011221
      \\[-1ex]
      \smallESixAffineLabels1021211
      \smallESixAffineLabels1112210
      \smallESixAffineLabels1121210
    \end{tabular}
    & $0$&$0$  & $16$ & $15$ & $0$ \\
  \end{tabular}
  \hspace{-2cm}
  \caption[$G_2$-weights of curves in the $IV^*$ Kodaira fiber with
  $\mathfrak{S}_3$ monodromy.]{The $\Gamma=\mathfrak{S}_3$-orbits of the $72$ curves, $G_2$ weight, genus
    of the moduli space, and the resulting number of hyper and vector
    multiplets. The weight under $G_2$ is the intersection product
    with 
    $C_1+C_2+C_3$ and $3C_0$.}
  \label{tab:E6G2matter}
\end{table}

For any gauge theory to arise, one needs to contract some of the fiber
irreducible components. Since there is no section, the standard
prescription cannot be applied. The minimal generalization would be to
contract all three irreducible components $V(w_2)\cap F$, and this is
what we will consider in the remainder of this subsection. This
corresponds to the $E_6\supset G_2$ branching rule, the composition of
$E_6\supset D_4$ from eq.~\eqref{eq:E6branchD4} with the
symmetric-type $D_4\supset G_2$ branching rule. There is no change in
the Weierstrass model compared to the non-split $E_6$, so we still
expect $14$ branch points over the $IV^*$ discriminant locus. This is
confirmed by a numerical analysis of a generic Calabi-Yau
hypersurface. There are three possible stabilizers under the symmetric
group action, leading to the following types of roots:
\begin{itemize}
\item 
  \begin{minipage}[t]{0.65\textwidth}
    The moduli space of a $\mathfrak{S}_3$-symmetric curve is still $\IP^1$. For
    example, the moduli space of the central fiber component $C_0 =
    V(w_0)\cap F$ is still $\IP^1$.
  \end{minipage}
  \hfill
  \raisebox{-0.5\height}{
    \ESixAffineLabels0000100
  }
\item 
  \begin{minipage}[t]{0.65\textwidth}
    The moduli space of a $\Z_2$-invariant curve is a triple cover of
    $\IP^1$ branched at $14$ points. Such a Riemann surface has genus
    $g'=5$. For example, $C_1$ which is one of the three irreducible
    components of $V(w_1)\cap F$.
  \end{minipage}
  \hfill
  \raisebox{-6mm}{
    \ESixAffineLabels0010000
  }
\item
  \begin{minipage}[t]{0.65\textwidth}
    The moduli space of a curve with trivial $\mathfrak{S}_3$-stabilizier is a
    sixfold cover over $\IP^1$ branched at $14$ points. At each
    ramification point, the six sheets meet in three pairs. Hence, the
    moduli space is a Riemann surface of genus $g'=16$.
  \end{minipage}
  \hfill
  \raisebox{-6mm}{
    \ESixAffineLabels1011221
  }
\end{itemize}
The spectrum is tabulated in \autoref{tab:E6G2matter} using the method
explained in \autoref{sec:other_end}. Reassembling the massless fields
into $G_2$ representations, we find one adjoint ($\Rep{14}$) vector
multiplet and $5+4+4=13$ hypermultiplets transforming in the the
fundamental $\Rep{7}$. There are also $15+15=30$ additional uncharged
hypermultiplets. To summarize, the generic Calabi-Yau hypersurface
$Y_3\subset X_{\nabla_3}$ has Hodge numbers $h^{1,1}(Y_3)=4$,
$h^{2,1}(Y_3)=70$. The 6-d F-theory compactification is a $G_2$ gauge
theory with $13\times \Rep{7}$ matter hypermultiplets. This satisfies
the gauge and gravity anomaly cancellation conditions\footnote{Similar
  to eq.~\eqref{eq:F4anomaly}, note that the charged dimension of
  $\Rep{7}$ is $6$.}
\begin{equation}
  \begin{gathered}
    H_u = h^{2,1} + 1 - 13\cdot 1 + 30 + 108 = 196
    ,\quad 
    H_c = 13 \cdot 7
    ,\quad V = 14
    \\
    H_u+H_c-V = 273 
    ,\quad 
    18 b = 13\cdot 2 - 8 
    ,\quad 
    3 b^2 = 13\cdot 1 - 10
  \end{gathered}
\end{equation}
for $b=1$.

\subsection{Alternative Limits}
\label{sec:E6toA2}

There is no particular reason to pick the monodromy orbit at the ends
of the affine $E_6$ Dynkin diagram as the curves to keep at finite
size. Certainly, there is no given section that would single out this
choice. Another possibility is to select the central
(monodromy-invariant) node. As we will see, this choice leads to a 5-d
theory with some interesting features, and suggests similar features
of the 6-d F-theory limit.

First, the unbroken gauge group is $A_2=SU(3)$: Not shrinking the
central node yields an $A_2^3$ gauge group as discussed in
\autoref{sec:D4andA23}. The $\mathfrak{S}_3$ monodromy breaks it further to the
diagonal $A_2$. Note that the monodromy is the same as in
\autoref{sec:3x_monodromy}, only the choice of shrinking curves
differs. Breaking the anomaly virtual representation by this branching
rule leads to
\begin{equation}
  H-V = 
  7 \times \Rep{27} \ominus \Ad(E_6)
  =
  14 \times \Ad\big(SU(3)\big) \oplus 19 \times \Rep{1}
  \ominus \Sym^3(\Rep{3}) \ominus \Sym^3(\barRep{3})
  .
\end{equation}
Surprisingly, this representation content is at odds with Witten's
rule~\cite{Witten:1996qb} for counting fields by the genus of the
moduli space of the curves (as reviewed in
\autoref{sec:other_end}). In particular, the na\"{\i}ve application would
result in the same $SU(3)$-representations for vectors and hypers in
contrast to the branching rule. The resolution of this puzzle must be
that there are additional massless degrees of freedom, and indeed
there ought to be: The central $\CP^1$ fiber component, which we chose
to leave at finite size, has multiplicity three. Therefore, one has
additional infinitesimal\footnote{They do not extend to finite
  deformations, which means that there is a superpotential term that
  prevents them from acquiring a vev.} deformations that ought to
manifest themselves as massless fields. 

We can understand this multiplicity in more detail by writing down the
projection map. For the smooth Calabi-Yau hypersurface, this is
\begin{equation}
  X_{\Delta_3} \to \CP^2, \quad
  [u:v:w_0:w_1:w_2:f_1:f_2:f_3] \mapsto [u:v:w_0^3 w_1^2 w_2] 
\end{equation}
for any smooth triangulation of the polytope
eq.~\eqref{eq:E6_3x_polytope} respecting the fibration structure. Now,
contracting all fiber components except the central node amounts to
removing the rays corresponding to $w_1$ and $w_2$ from the fan and
merging their star into a single cone, making the triangulation
coarser. The projection map of this singular toric variety is
\begin{equation}
  X_{\Delta_3'} \to \CP^2, \quad
  [u:v:w_0:f_1:f_2:f_3] \mapsto [u:v:w_0^3] 
\end{equation}
and we see that turning on $w_0$ infinitesimally does indeed not move
the fiber to first order. We note that this is a novel feature of not
having a section: A section can only intersect the fiber in a single
point (counted with multiplicity), that is, a point on a fiber
component with multiplicity one. Only a three-section can intersect
the central node of the affine $E_6$ in a single point of multiplicity
three. 
\begin{table}[htbp]
  \centering
  \begin{tabular}{c@{$\quad ($}r@{,\;}r@{$)\quad$}cccc}
    % shrink and symmetric
    \multicolumn{1}{c}{$\Gamma$-orbit} & 
    \multicolumn{2}{c}{$A_2$-wt.} & 
    $g'$ & H & V
    \\ \hline
    % S3-symmetric
    \smallESixAffineLabels1111211
    & $-3$&$3$  & $0$ & $0$ & $1$ \\
    \smallESixAffineLabels0000100
    & $-3$&$0$  & $0$ & $0$ & $1$ \\
    \smallESixAffineLabels1111111
    & $0$&$3$   & $0$ & $0$ & $1$ \\
    \smallESixAffineLabels0011210
    & $0$&$-3$  & $0$ & $0$ & $1$ \\
    \smallESixAffineLabels1122221
    & $3$&$0$   & $0$ & $0$ & $1$ \\
    \smallESixAffineLabels0011110
    & $3$&$-3$  & $0$ & $0$ & $1$ \\
    % Z2-symmetric
    \smallESixAffineLabels1112321
    \smallESixAffineLabels1121321
    \smallESixAffineLabels1122311
    & $-2$&$1$  & $5$ & $6$ & $2$ \\
    \smallESixAffineLabels1111221
    \smallESixAffineLabels1112211
    \smallESixAffineLabels1121211
    & $-1$&$2$  & $5$ & $5$ & $1$ \\
    \smallESixAffineLabels0000110
    \smallESixAffineLabels0001100
    \smallESixAffineLabels0010100
    & $-1$&$-1$  & $5$ & $5$ & $1$ \\
    \smallESixAffineLabels1112221
    \smallESixAffineLabels1121221
    \smallESixAffineLabels1122211
    & $1$&$1$  & $5$ & $5$ & $1$ \\
    \smallESixAffineLabels0001110
    \smallESixAffineLabels0010110
    \smallESixAffineLabels0011100
    & $1$&$-2$  & $5$ & $5$ & $1$ \\
    \smallESixAffineLabels0000010
    \smallESixAffineLabels0001000
    \smallESixAffineLabels0010000
    & $2$&$-1$  & $5$ & $6$ & $2$ \\
    % finite size and Z2-symmetric
    \smallESixAffineLabels0000111
    \smallESixAffineLabels0101100
    \smallESixAffineLabels1010100
    & $-2$&$1$  & $5$ & $4$ & $0$ \\
    \smallESixAffineLabels0111211
    \smallESixAffineLabels1011211
    \smallESixAffineLabels1111210
    & $-2$&$1$  & $5$ & $4$ & $0$ \\
    \smallESixAffineLabels0000001
    \smallESixAffineLabels0100000
    \smallESixAffineLabels1000000
    & $-1$&$2$  & $5$ & $5$ & $1$ \\
    \smallESixAffineLabels0101111
    \smallESixAffineLabels1010111
    \smallESixAffineLabels1111100
    & $-1$&$2$  & $5$ & $4$ & $0$ \\
    \smallESixAffineLabels0011211
    \smallESixAffineLabels0111210
    \smallESixAffineLabels1011210
    & $-1$&$-1$  & $5$ & $4$ & $0$ \\
    \smallESixAffineLabels0112321
    \smallESixAffineLabels1021321
    \smallESixAffineLabels1122310
    & $-1$&$-1$  & $5$ & $5$ & $1$ \\
    \smallESixAffineLabels0000011
    \smallESixAffineLabels0101000
    \smallESixAffineLabels1010000
    & $1$&$1$  & $5$ & $5$ & $1$ \\
    \smallESixAffineLabels0111111
    \smallESixAffineLabels1011111
    \smallESixAffineLabels1111110
    & $1$&$1$  & $5$ & $4$ & $0$ \\
    \smallESixAffineLabels0011221
    \smallESixAffineLabels0112210
    \smallESixAffineLabels1021210
    & $1$&$-2$  & $5$ & $4$ & $0$ \\
    \smallESixAffineLabels0122321
    \smallESixAffineLabels1022321
    \smallESixAffineLabels1122320
    & $1$&$-2$  & $5$ & $5$ & $1$ \\
    \smallESixAffineLabels0011111
    \smallESixAffineLabels0111110
    \smallESixAffineLabels1011110
    & $2$&$-1$  & $5$ & $4$ & $0$ \\
    \smallESixAffineLabels0112221
    \smallESixAffineLabels1021221
    \smallESixAffineLabels1122210
    & $2$&$-1$  & $5$ & $4$ & $0$
    % finite size and nonsymmetric
    \\[1ex]
    \begin{tabular}{c}
      \smallESixAffineLabels0001111
      \smallESixAffineLabels0010111
      \smallESixAffineLabels0101110
      \\[-1ex]
      \smallESixAffineLabels0111100
      \smallESixAffineLabels1010110
      \smallESixAffineLabels1011100
    \end{tabular}
    & $0$&$0$  & $16$ & $15$ & $0$ 
    \\[2ex]
    \begin{tabular}{c}
      \smallESixAffineLabels0111221
      \smallESixAffineLabels0112211
      \smallESixAffineLabels1011221
      \\[-1ex]
      \smallESixAffineLabels1021211
      \smallESixAffineLabels1112210
      \smallESixAffineLabels1121210
    \end{tabular}
    & $0$&$0$  & $16$ & $15$ & $0$ \\
  \end{tabular}
  \hspace{-2cm}
   \caption[$A_2$-weights of curves in the $IV^*$ Kodaira fiber with
   $\mathfrak{S}_3$ monodromy.]
   {The $\Gamma=\mathfrak{S}_3$-orbits of the $72$ curves, $A_2$ weight, genus
     of the moduli space, and the resulting number of hyper and vector
     multiplets. The weight under $A_2$ is the intersection product with
     $C_1+C_2+C_3$ and $C_4+C_5+C_6$.}
   \label{tab:E6A2matter}
\end{table}
Hence, we propose the following addition to the Witten's rule for
counting the massless spectrum:
\begin{itemize}
\item Each curve $C$ that does not wrap a fiber irreducible component
  whose multiplicity is one yields an additional hyper/vector
  multiplet pair.
\end{itemize}
Note that, as before, this is meant to apply to curves such that
either $C$ or $F-C$ does not wrap a multiplicity-one curve, that is,
either $a_4=a_5=a_6=0$ or $a_4=a_5=a_6=1$. In \autoref{tab:E6A2matter}
we applied this modified rule to count the hyper- and vector
multiplets.

In terms of $A_2$-representations, the weights na\"{\i}vely assemble into
$\Sym^3(\Rep{3})\oplus \Sym^3(\barRep{3})$ of vector multiplets and
vector and $14$ adjoint hypermultiplets, which is at first sight quite
an odd field content. However, also note that the weights live in an
index-$3$ sublattice of the $A_2$ weight lattice. Geometrically, this
is because $(C_1+C_2+C_3)-(C_4+C_5+C_6)$ is divisible by $3$ modulo
$F=3C_0+2(C_1+C_2+C_3)+(C_4+C_5+C_6)$. Hence, we should have used
these refined curve classes which amounts to the intersection numbers
spanning the whole weight lattice instead of a index-$3$
sublattice. Hence, we identify the actual field content as vector
multiplets transforming in the $\Ad\big(SU(3))\oplus 2\times
(\Rep{3}\oplus \barRep{3})$ and hypermultiplets transforming as
$14\times(\Rep{3}\oplus\barRep{3})$. The vector multiplets not in the
adjoint can pair up with hypermultiplets and become massive, leaving
us with an ordinary $SU(3)$ gauge theory with
$12\times(\Rep{3}\oplus\barRep{3})$ charged hypermultiplets, canceling
the non-Abelian gauge anomaly
\begin{equation}
  18 b = 12\cdot (1+1) - 6
  ,\quad 
  3 b^2 = 12\cdot (\tfrac{1}{2} + \tfrac{1}{2}) - 9
\end{equation}
for $b=1$. To count the uncharged hypermultiplets, note that there are
again $2\times 15$ uncharged fields localized along the discriminant, see
\autoref{tab:E6A2matter}, and $108$ fields localized in
codimension-two at the cusps of the $I_1$ discriminant
component. Together, they cancel the gravitational anomaly as
\begin{equation}
  H_u = h^{2,1} + 1 + 30 + 108
  ,\quad 
  H_c = 12 \cdot (3+3)
  ,\quad
  V = 8
  ,\quad
  H_u+H_c-V = 273 
  .
\end{equation}

%%% Local Variables:
%%% TeX-master: "Main.tex"
%%% eval: (TeX-PDF-mode 1)
%%% End:

\section{The Shioda-Tate(-Wazir) Formula}
\label{sec:STW}

The divisors in a smooth elliptic fibration $\pi:Y\to B$ are either
(multi-)sections, that is, map to all of $B$, or vertical, that is,
their image is again codimension-one in the base. The vertical case
can be further subdivided in the case where divisor contains the
entire fiber or only irreducible fiber components. In particular, one
has the following divisors:
\begin{itemize}
\item The given section $\sigma$,
\item Sections $S_i$, $i = 1,\dots,\rank MW(Y)$ forming a basis of the
  Mordell-Weil group,
\item Pull-backs from the base, $B_j = \pi^{-1}(b_j)$,
  $j=1,\dots,h^{1,1}(Y)$,
\item and \emph{fibral divisors} $T_{\delta, k}$ that are irreducible
  components of the preimage $\pi^{-1}(\delta)$ of irreducible
  components of the discriminant, $\delta \subset \Delta$. The
  subscript $k = 0,\dots,K_\delta-1$ labels the irreducible fiber
  components modulo the monodromy action.
\end{itemize}
To avoid obvious homology relations, we considered a basis for the
Mordell-Weil group and the base divisors $b_j$ here. For the fibral
divisors $T_{\delta,k}$ one further notes that the sum over all fiber
components is already generated by the $B_j$, so we have to exclude
one of the fiber components. Customarily, one excludes the component
intersecting the given section (labeled $k=0$), so we demand that
$k\geq 1$. 
The Shioda-Tate formula \cite{MR0429918,MR0225778,MR1610977}
(extended to threefolds by Wazir \cite{2001math.....12259W})
statess
that these generate all divisor classes, that is,
\begin{equation}
  h^{1,1}(Y) = 
  1 + \rank MW(Y) + h^{1,1}(B) + \sum_\delta (K_\delta -1)
\end{equation}

Now consider a genus-one fibration $Y'\to B$ without a section. Even
if there is no section, there is always some multi-section $\sigma'$
for a sufficiently high degree for a \emph{projective} genus-one
fibration, generating some one-dimensional subspace of
$h^{1,1}(Y')$. Hence, the generalization of Shioda-Tate-Wazir formula
still contains the ``1+'' part. 

As for the fibral divisors, the story also generalizes in an obvious
way. Even if there are more general monodromies, one still has to
group irreducible fiber components into the $K_\delta'$ monodromy
orbits. One of these fibral divisors needs to be excluded to avoid
homology relations with the divisors $B_j$. We cannot rely on a
section to pick it, but any choice is fine.

Finally, consider the term involving the Mordell-Weil group. For a
genus-one fibration $Y'$ we still have an associated Mordell-Weil
group $MW(J(Y'))$ of the Jacobian. Recall that the Mordell-Weil group
acts by translations (birationally) on the points of the elliptic
fibration. This action commutes with the twist by an element of the
Tate-Shafarevich group, and therefore extends to an action on the
genus-one fibration. Hence we can act on the chosen multi-section
$\sigma'$ to produce new multi-sections $S_i'(\sigma')$ of the same
degree. Moreover, free generators in the Mordell-Weil group cannot fix
$\sigma'$. In fact, the set of images $S_i'(\sigma')$ for a basis
$i=1, \dots, \rank MW(J(Y'))$ are again independent multi-sections in
homology.

To summarize, the Shioda-Tate-Wazir formula generalized to
\begin{equation}
  h^{1,1}(Y') =
  1 + \rank MW(J(Y')) 
  + h^{1,1}(B) + \sum_\delta (K_\delta' -1)
\end{equation}
for genus-one fibrations. By the analogous argument as in the case
with a section, we can identify the Mordell-Weil lattice with the
Abelian gauge bosons in the F-theory compactification. That is, the
F-theory gauge group is $G \times U(1)^r$ with $r=\rank MW(J(Y))$ and
$G$ a simple Lie group.

%%% Local Variables:
%%% TeX-master: "Main.tex"
%%% eval: (TeX-PDF-mode 1)
%%% End:

\appendix

\section{Representation Theory in Sage}
\label{sec:sage}

The representation theory of semisimple Lie groups is of course a
well-known subject. However the richness of representations and
branchings often requires us to look through tables, which is quite
tedious. In this appendix we would like to give a quick introduction
to Sage~\cite{Sage, Sage-Combinat} (\url{http://www.sagemath.org}) and
show how one can use it to compute with representations. Let us start
by defining the adjoint representation of $E_6$:
\begin{lstlisting}[style=SageInput]
sage: E6 = WeylCharacterRing('E6', style='coroots')
sage: E6.dynkin_diagram()
\end{lstlisting}
\begin{lstlisting}[style=SageOutput]
        O 2
        |
        |
O---O---O---O---O
1   3   4   5   6
E6
\end{lstlisting}
\begin{lstlisting}[style=SageInput]
sage: AdjE6 = E6(0,1,0,0,0,0)
sage: AdjE6              # print the representation
\end{lstlisting}
\begin{lstlisting}[style=SageOutput]
E6(0,1,0,0,0,0)
\end{lstlisting}
\begin{lstlisting}[style=SageInput]
sage: AdjE6.degree()     # its dimension
\end{lstlisting}
\begin{lstlisting}[style=SageOutput]
78
\end{lstlisting}
\begin{lstlisting}[style=SageInput]
sage: AdjE6 * AdjE6      # tensor product
\end{lstlisting}
\begin{lstlisting}[style=SageOutput]
E6(0,0,0,0,0,0) + E6(0,1,0,0,0,0) + E6(0,2,0,0,0,0) + 
E6(0,0,0,1,0,0) + E6(1,0,0,0,0,1)
\end{lstlisting}
Here, we used the \emph{coroots} (a.k.a.\ Dynkin labels), which are
non-negative integers corresponding to the nodes of the Dynkin
diagram, to define the representation \textsage{AdjE6}. The
\textsage{branch()} method computes the branching of the
representation to a subgroup. In simple cases it will be able to guess
the desired branching rule, for example the Levi-type branching rule
associated to removing a node from the Dynkin diagram:
\begin{lstlisting}[style=SageInput]
sage: D5 = WeylCharacterRing('D5', style='coroots')
sage: AdjE6.branch(D5)    # defaults to rule='levi'
\end{lstlisting}
\begin{lstlisting}[style=SageOutput]
D5(0,0,0,0,0) + D5(0,0,0,1,0) + D5(0,0,0,0,1) + D5(0,1,0,0,0)
\end{lstlisting}
However, in general you have to specify it with the
\textsage{rule=<rulename>} keyword option. For example, applying the
Levi-type rule twice:\footnote{Requires Sage version 6.1 or later.}
\begin{lstlisting}[style=SageInput]
sage: D4 = WeylCharacterRing('D4', style='coroots')
sage: levi2x = branching_rule(E6, D5, rule='levi') * \
....:          branching_rule(D5, D4, rule='levi')
sage: levi2x
\end{lstlisting}
\begin{lstlisting}[style=SageOutput]
composite branching rule E6 => (levi) D5 => (levi) D4
\end{lstlisting}
\begin{lstlisting}[style=SageInput]
sage: AdjE6.branch(D4, rule=levi2x)
\end{lstlisting}
\begin{lstlisting}[style=SageOutput]
2*D4(0,0,0,0) + 2*D4(0,0,1,0) + 2*D4(0,0,0,1) + 
2*D4(1,0,0,0) + D4(0,1,0,0)
\end{lstlisting}
In addition to the Levi-type branching rules, there is a variety of
other ones implemented. In particular, all branchings to maximal
subgroups are. As a fancy example, here is the (non-maximal) branching
rule associated to the automorphism of the $D_4$ Dynkin diagram:
\begin{lstlisting}[style=SageInput]
sage: G2 = WeylCharacterRing('G2', style='coroots')
sage: D4(0,1,0,0).branch(G2, rule='symmetric')
\end{lstlisting}
\begin{lstlisting}[style=SageOutput]
2*G2(1,0) + G2(0,1)
\end{lstlisting}
\begin{lstlisting}[style=SageInput]
sage: D4(0,1,0,0).degree(), G2(1,0).degree(), G2(0,1).degree()
\end{lstlisting}
\begin{lstlisting}[style=SageOutput]
(28, 7, 14)
\end{lstlisting}
So the irreducible representations decompose as $\Ad(D_4) = 2\times
\Rep{7} \oplus \Rep{14}$ for this particular embedding $D_4\supset
G_2$. Each representation can be restricted to a maximal torus
$T\subset G$, where it decomposes further into a sum of
one-dimensional representations since $T$ is Abelian. These are the
weights of the representation, and can be enumerated using
\begin{lstlisting}[style=SageInput]
sage: G2(1,0).weight_multiplicities()
\end{lstlisting}
\begin{lstlisting}[style=SageOutput]
{(-1, 1, 0): 1, (0, 1, -1): 1, (1, 0, -1): 1, (0, 0, 0): 1, 
 (-1, 0, 1): 1, (0, -1, 1): 1, (1, -1, 0): 1}
\end{lstlisting}
In particular, we can read off the multiplicity of the trivial weight
having multiplicity $1$ in $\Rep{7}$. The number of all non-trivial
weights is the \emph{charged dimension}~\cite{Grassi:2000we,
  Grassi:2011hq}, and we just computed that the charged dimension of
$\Rep{7}$ is $6$.

Finally, one would like to know the anomaly coefficients. These are
integers $A_R$, $B_R$, and $C_R$ associated to representations $R$
such that\footnote{If the group $G$ does not have an independent
  quartic Casimir operator then the two parameters $B_R$, $C_R$ are
  not uniquely determined. It is customary to define $B_R=0$ in that
  case. The computation of the anomaly coefficients is analogous but
  simpler, and it suffices to just consider a $SU(2)\subset G$
  subgroup instead of $SU(4)$.}
\begin{equation}
  \label{eq:anomaly_coeffs}
  \tr_R F^2 = A_R \tfrac{1}{\lambda_G}\tr F^2
  ,\quad
  \tr_R F^4 = 
  B_R \tfrac{1}{\lambda_G} \tr F^4 + 
  C_R \big( \tfrac{1}{\lambda_G} \tr F^2 \big)^2
  ,
\end{equation}
where $\tr_R$ is the trace over the Lie algebra generator in the
representation $R$ and $\tr$ is the trace over the fundamental
representation. We will be using the \emph{integral
  normalization}~\cite{Grassi:2011hq} where the factors $\lambda_G$
listed in \autoref{tab:lambda} are absorbed into the anomaly
coefficients. That is, we use the properly normalized
$\tfrac{1}{\lambda_G} \tr$ for the trace over the fundamental
representation in eq.~\eqref{eq:anomaly_coeffs}. However, note that
much of the physics literature uses the convention where the
$\lambda_G$ prefactors are not absorbed.
\begin{table}[htbp]
  \centering
  \begin{math}
    \displaystyle
  \begin{array}{c|ccccccccc}
    G & A_n & B_n & C_n & D_n & G_2 & F_4 & E_6 & E_7 & E_8 
    \\\hline
    \lambda_G & 
    1 & 2 & 1 & 2 & 
    2 & 6 & 6 & 12 & 60
  \end{array}
  \end{math}
  \caption{Extra normalization factor in the fundamental trace.}
  \label{tab:lambda}
\end{table}
This then requires compensating factors of $\lambda_G$ in the gauge
anomaly cancellation condition, which is why we are not following this
convention. The traces on the right hand sides of
eq.~\eqref{eq:anomaly_coeffs} are easily determined by restriction,
that is, branching to a subgroup. We will be using $SU(4)$, the
simplest group with a quartic Casimir. There are various bases one can
use for the representations of $SU(4)$, for example
\begin{itemize}
\item Irreducible representation $R(i,j,k)$ indexed by coroots,
\item Tensor products $\Rep{4}^i \otimes \Rep{6}^j \otimes
  \barRep{4}^k$, and
\item Tensor products of symmetrizations 
  \begin{equation}
    S(i,j,k) \eqdef
    \Sym^i(\Rep{4}) \otimes
    \Sym^j(\Rep{6}) \otimes 
    \Sym^k(\barRep{4}) 
    .
  \end{equation}
\end{itemize}
In either basis one can explicitly construct representations in terms
of tensor operations. The traces for irreducible representations of
the subgroup can then be evaluated directly. The third basis has the
advantage that, for the two particular Lie algebra generators
\begin{equation}
  F_1 = \frac{1}{\sqrt{2}} \diag(+1,-1,0,0)
  ,\quad
  F_2 = \frac{1}{\sqrt{2}} \diag(0,0,+1,-1)
\end{equation}
of $SU(4)$, one can explicitly evaluate the traces as polynomials in
$i$, $j$, and $k$. They are
\begin{equation}
  \begin{split}
    \tr_{S(i,j,k)}& F^2 =
    \frac{1}{1814400}
    \prod_{x=1}^3 (i + x)
    \prod_{y=1}^5 (j + y)
    \prod_{z=1}^3 (k + z)
    \\ 
    & \quad
    \times
    \big(
    21 i^2 + 20 j^2 + 21 k^2 + 84 i + 120 j + 84 k
    \big)
    ,
     \\
    \tr_{S(i,j,k)}& F^4 = 
    \frac{1}{1814400}
    \prod_{x=1}^3 (i + x)
    \prod_{y=1}^5 (j + y)
    \prod_{z=1}^3 (k + z)
    \\ & \hspace{-1cm}
    \times
    \big(
    30 i^4 + 60 i^2 j^2 + 25 j^4 + 63 i^2 k^2 + 60 j^2 k^2 
    + 30 k^4 
    + 240 i^3
    + 360 i^2 j 
    \mathrlap{+ 240 i j^2}
    \\ & \hspace{-5mm}
    + 300 j^3 
    + 252 i^2 k + 240 j^2 k + 252 i k^2 + 360 j k^2 + 240 k^3 
    + 435 i^2 
    \\ & \hspace{-5mm}
    + 1440 i j + 825 j^2 + 1008 i k + 1440 j k 
    + 435 k^2 - 180 i - 450 j - 180 k 
    \big)
    ,
    \\
    \tr_{S(i,j,k)}& F_1^2 F_2^2 =
    \frac{1}{163296000}
    \prod_{x=1}^3 (i + x)
    \prod_{y=1}^5 (j + y)
    \prod_{z=1}^3 (k + z)
    \\ & \hspace{-1cm}
    \times
    \big(
    45 i^4 + 180 i^2 j^2 + 125 j^4 + 189 i^2 k^2 + 180 j^2 k^2 
    + 45 k^4 + 360 i^3 + 1080 i^2 j
    \\ & \hspace{-5mm}
    + 720 i j^2 + 1500 j^3 
    + 756 i^2 k + 720 j^2 k + 756 i k^2 + 1080 j k^2 + 360 k^3 
    \\ & \hspace{-5mm}
    + 495 i^2 + 4320 i j + 4525 j^2 + 3024 i k + 4320 j k 
    + 495 k^2 - 900 i + 150 j 
    \mathrlap{- 900 k \big).}
  \end{split}
\end{equation}
Since the (usually reducible) representations $S(i,j,k)$ form a basis,
one can express any representation as a linear combination of
them. The anomaly coefficients then are
\begin{equation}
  \begin{split}
    A_R \;&= 
    \lambda_G \frac{\tr_R F_1^2}{\tr F_1^2} = 
    \lambda_G \frac{\tr_R F_2^2}{\tr F_2^2} \\
    B_R \;&= 
    \lambda_G
    \frac{ 
      \tr_R F_1^4 - 3 \tr_R F_1^2 F_2^2
    }{
      \tr F_1^4
    }    
    =
    \lambda_G
    \frac{ 
      \tr_R F_2^4 - 3 \tr_R F_1^2 F_2^2
    }{
      \tr F_2^4
    }    
    \\
    C_R \;&= 
    \lambda_G^2 
    \frac{
      3 \tr_R F_1^2 F_2^2
    }{
      \big( \tr F_1^2 \big) \big( \tr F_2^2 \big)
    }
  \end{split}
\end{equation}

We implemented the above algorithm in Sage, which one can find and use
at the following URL:
\begin{lstlisting}[style=SageInput, breaklines=true]
sage: load('http://boxen.math.washington.edu/home/vbraun/www/anomaly_coefficients.py')
sage: E6 = WeylCharacterRing('E6', style='coroots')
sage: anomaly_coefficients(E6(1,0,0,0,0,0))  # the 27 of E6
\end{lstlisting}
\begin{lstlisting}[style=SageOutput]
{'A': 6, 'B': 0, 'C': 3}
\end{lstlisting}
\begin{lstlisting}[style=SageInput]
sage: E6(3,2,0,0,0,0).degree()
\end{lstlisting}
\begin{lstlisting}[style=SageOutput]
3162159
\end{lstlisting}
\begin{lstlisting}[style=SageInput]
sage: anomaly_coefficients(E6(3,2,0,0,0,0))  # about 30 minutes
\end{lstlisting}
\begin{lstlisting}[style=SageOutput]
{'A': 5027022, 'C': 22621599, 'B': 0}
\end{lstlisting}
The group $D_4 = Spin(8)$ is special in that it has three independent
Casimir operators in degree $4$. There are two ways to handle this,
either by introducing an additional anomaly coefficient and another
gauge anomaly cancellation condition or by demanding that the
$F^4$-anomaly still cancels if one acts by triality. Note that the
$B_R$, $C_R$ anomaly coefficients transform non-trivially under triality:
\begin{lstlisting}[style=SageInput]
sage: D4 = WeylCharacterRing('D4', style='coroots')
sage: D4.dynkin_diagram()
\end{lstlisting}
\begin{lstlisting}[style=SageOutput]
    O 4
    |
    |
O---O---O
1   2   3
D4
\end{lstlisting}
\begin{lstlisting}[style=SageInput]
sage: anomaly_coefficients(D4(1,0,0,0))   # 8_spinor
\end{lstlisting}
\begin{lstlisting}[style=SageOutput]
{'A': 2, 'C': 0, 'B': 2}
\end{lstlisting}
\begin{lstlisting}[style=SageInput]
sage: anomaly_coefficients(D4(0,0,1,0))   # 8_conjugate
\end{lstlisting}
\begin{lstlisting}[style=SageOutput]
{'A': 2, 'C': 0, 'B': 2}
\end{lstlisting}
\begin{lstlisting}[style=SageInput]
sage: anomaly_coefficients(D4(0,0,0,1))   # 8_vector
\end{lstlisting}
\begin{lstlisting}[style=SageOutput]
{'A': 2, 'C': 3, 'B': -4}
\end{lstlisting}
As a final example, consider the $SU(N)$-representation
{\tiny$\yng(2,2)$}, that is, with coroots $(0,2,0,\dots,0)$. Knowing
that the anomaly coefficients are polynomials in $N$, we can easily
determine them (see also Table~1 of~\cite{Kumar:2010am}):
\begin{lstlisting}[style=SageInput]
sage: def Young2x2_SU(N):
....:     SU_N = WeylCharacterRing(('A',N-1), style='coroots')
....:     young_box_2x2 = SU_N(0,2)
....:     return anomaly_coefficients(young_box_2x2)
sage: R.<N> = QQ[]
sage: for X in ['A', 'B', 'C']:
....:     data = [(N, Young2x2_SU(N)[X]) for N in range(4,10)]
....:     print X, ':' , R.lagrange_polynomial(data).factor()
\end{lstlisting}
\begin{lstlisting}[style=SageOutput]
A : (1/3) * (N - 2) * N * (N + 2)
B : (1/3) * N * (N^2 - 58)
C : (3) * (N^2 + 2)
\end{lstlisting}

%%% Local Variables:
%%% TeX-master: "Main.tex"
%%% eval: (TeX-PDF-mode 1)
%%% End:

\section*{Acknowledgments}

We thank 
Lara Anderson, 
Paul Aspinwall,
Andreas Braun,
Andr\'es Collinucci,
Daniel Bump,
Antonella Grassi,
Simeon Hellerman,
Kentaro Hori,
Gregory Moore,
Nathan Seiberg,
Washington Taylor,
and
Edward Witten
for discussions.
We gratefully acknowledge the hospitality of the Simons Center for
Geometry and Physics as well as the University of Pennsylvania where
part of this work was performed; D.R.M.\ also thanks the Aspen Center
for Physics for hospitality.  The work of D.R.M.\ was supported in
part by the National Science Foundation under grants DMS-1007414, PHY-1066293,
and PHY-1307513.  The work of V.B.\ was supported in part by the
Dublin Institute for Advanced Studies.

\bibliographystyle{utcaps} 
\renewcommand{\refname}{Bibliography}
\addcontentsline{toc}{section}{Bibliography} 
\bibliography{Main}

\end{document}